\documentclass[journal=jpccck,manuscript=article]{achemso}

\usepackage[version=3]{mhchem} 
\usepackage{graphicx}
\usepackage{amsmath, amssymb}
\usepackage{subcaption}
\usepackage{orcidlink}
\usepackage{docmute}

\title{Size asymmetry and dynamic cluster persistence in quasi-two-dimensional colloids with competitive interactions}

\author{Chieh-Chih George Yeh,\orcidlink{0000-0001-9302-6194}}
\affiliation[UTChE]{McKetta Department of Chemical Engineering, University of Texas at Austin, Austin, TX, USA}
\alsoaffiliation[UMChE]{Department of Chemical Engineering, University of Michigan, Ann Arbor, MI, USA}
\author{Jin Gyun Lee,\orcidlink{0000-0002-5708-2284}}
\affiliation[LSUChE]{Cain Department of Chemical Engineering, Louisiana State University, Baton Rouge, LA, USA}
\author{Paul Bogdan,\orcidlink{0000-0003-2118-0816}}
\affiliation[USC]{Viterbi School of Engineering, University of Southern California, Los Angeles, California 90089, USA}
\alsoaffiliation[UMC]{Center for Complex Particle Systems (COMPASS), University of Michigan, Ann Arbor, Michigan 48109, USA}
\author{Bhuvnesh Bharti,\orcidlink{0000-0001-9426-9606}}
\affiliation[LSUChE]{Cain Department of Chemical Engineering, Louisiana State University, Baton Rouge, LA, USA}
\email{bbharti@lsu.edu}
\author{Thomas M. Truskett,\orcidlink{0000-0002-6607-6468}}
\affiliation[UTChE]{Department of Chemical Engineering, University of Michigan, Ann Arbor, MI, USA}
\alsoaffiliation[UMC]{Center for Complex Particle Systems (COMPASS), University of Michigan, Ann Arbor, MI, USA}
\alsoaffiliation[UMB]{Biointerfaces Institute, University of Michigan, Ann Arbor, MI, USA}
\alsoaffiliation[UTChE]{McKetta Department of Chemical Engineering, University of Texas at Austin, Austin, TX, USA}
\email{truskett@umich.edu}
\abbreviations{BD, MC, SALR}

\begin{document}
\maketitle

\begin{abstract}
Nonequilibrium assembly of colloids with short-range attractive and long-range repulsive interactions can produce a wide variety of microstructures, but predicting assembly outcomes remains challenging, limiting opportunities to engineer materials with targeted properties. Here, we use Brownian dynamics simulations and experiments to study the formation of colloidal clusters that emerge in quasi-two-dimensional mixtures of large and small nonmagnetic microsphere monolayers immersed in a magnetized ferrofluid. Application of a magnetic field normal to the monolayer allows for dynamic modulation of the competition between attractions and repulsions, and hence the emergent structure and dynamics, in a manner that depends on the microsphere sizes and mixing ratio. To characterize these complex assemblies, we develop and compare complementary structural and topological cluster descriptors: a geometric classifier that resolves local symmetry via identification of crystalline alloy motifs and a force-weighted graph analysis that quantifies Olivier--Ricci curvature. We find that dynamic persistence of clusters depends nonmonotonically on microsphere size ratio, but correlates with the emergence of close-packed large-particle motifs or binary alloy structures in which large-particle cages tightly enclose small microspheres to maximize attractions. We further show that characteristic persistence times strongly anticorrelate with Ollivier-Ricci curvature of force-weighted cluster graphs across the simulated dataset. Clusters dissociate at high magnetic fields, except for in microsphere mixtures with a 2:1 radius ratio, which projects to the magic radius ratio for the stable S1 square phase [$1:(\sqrt{2}-1)\approx 2.414$] in the quasi-two-dimensional monolayer.     
\end{abstract}

\section{Introduction}

Colloidal assembly offers a versatile platform for fabricating solution-processed, micro- or nanoparticle-based superstructures with collective properties not found in conventional materials\cite{lotito2017approaches,li2022colloidal,lalanne2025disordered}. Colloids that assemble due to competing short-range interparticle attractions (SA) and long-range repulsions (LR), termed SALR colloids, are particularly interesting candidates for material design because their interactions can drive formation of a wide variety of equilibrium and non-equilibrium assemblies, including cluster fluids, percolated gels, and glasses\cite{segre2001glasslike, sciortino2004equilibrium, stradner2004equilibrium, campbell2005dynamical,godfrin2014generalized,jadrich2015origin,schwanzer2016two,lindquist2016assembly,zhuang2016recent,das2017clustering,lindquist2017interactions,pham2017phase,saez2018gelation,perdomo2022thermodynamics,al2022dual,ruiz2021role}. As with spherical particles featuring other types of interactions\cite{likos1993complex,dijkstra1999phase,shevchenko2006structural,filion2009prediction,dong2011two,bonales2012phase,ye2015structural,steurer2016intermetallics,lacour2019influence,lotito2018pattern,fayen2020infinite,padilla2021dynamics,lopez2022self,fayen2024quasicrystal}, the assembly landscape for SALR colloids expands considerably if binary mixtures of particles with two sizes are considered.\cite{pandey2013dynamics,pandey2016gelation,toyotama2016spontaneous,yeh2026dynamic}.    
Size bidispersity breaks the favored translational and rotational symmetries of monodisperse colloids, leading to distinct structural motifs characteristic of binary superlattices, quasicrystals, or heterogeneous glassy networks.\cite{ho2006three, bonales2012phase,fayen2020infinite, padilla2021dynamics}. Mechanistically, varying the colloid size ratio modulates the interstitial void geometry and reshapes size-dependent SALR interactions.\cite{filion2009prediction,kozina2011polymer,hopkins2012densest,law2013self,parola2015depletion,fayen2024quasicrystal,yeh2026dynamic} Particle packing efficiency in particular shows pronounced non-monotonic dependency on colloid size ratio, reflecting specific geometries (magic radii ratios) where the smaller particles can just fit into interstitial holes between larger particles without lattice distortion,\cite{likos1993complex} increasing the load-bearing contacts and the maximum local packing fractions \cite{mcgeary1961mechanical, zheng1995packing}.

Here, we seek to understand how particle-size ratio can be leveraged to manipulate structure formation in dynamically tunable SALR (d-SALR) colloids. To realize d-SALR colloids, an experimental platform was recently introduced in which a nonmagnetic microsphere monolayer was immersed in a ferrofluid subject to a normally directed external magnetic field.\cite{al2022dual, gauri2023magnetic} The solvent-dispersed magnetic nanoparticles of the ferrofluid induced competitive dipolar repulsions and short-range, depletion attractions between the microspheres. While the depletion attractions were set by the microsphere and nanoparticle dimensions and concentrations, the magnetic-field-induced dipolar repulsions were tunable by the strength of the applied magnetic field. As a result, the preferred microsphere assembly could be modified on-the-fly by external field changes in a way that could be understood and modeled theoretically considering the balance of interparticle attractions and repulsions.\cite{al2022dual, gauri2023magnetic,yeh2025colloidal}. At low fields, depletion interactions between microspheres favored assembly of a close-packed solid. Increasing magnetic field strength caused the dipolar repulsions to balance the depletion attractions, stabilizing finite-size microsphere clusters. At sufficiently high fields, the dipolar repulsions overwhelmed the attractions, promoting assembly of open (Wigner) solids. 

Adding a second nonmagnetic microsphere to the monolayer with a radius half that of the first introduced nonadditivity and asymmetry to the pairwise d-SALR interactions, qualitatively changing the assembly behavior described above.\cite{yeh2026dynamic} The most prominent impact was the generation of nontrivial cross-interaction effects between the two species similar to those predicted by theoretical models.\cite{ferreiro2018cluster,tan2021cluster,tan2024secondary,patsahan2024spontaneous} Specifically, the balance between field-tunable dipolar interactions and depletion attractions was different for large-large, large-small, and small-small microsphere pairs, with cross-species attractions dominating over like-species attractions at higher magnetic-field strengths. In both experiments and Brownian dynamics (BD) simulations, these interactions favored compositionally mixed, nonequilibrium cluster states, which differed qualitatively from the phase-separated, species-enriched assemblies that formed at equilibrium. Furthermore, the dynamic persistence of the nonequilibrium clusters could be sensitively controlled via the mixing ratio of small-to-large microspheres and the field strength. The cluster exchange dynamics were shown to correlate with Ollivier-Ricci curvature (ORC) of force-weighted graphs constructed to quantify cluster interaction topology, providing a link between microscopic cluster organization and dynamics.  


In the present study, we build on these observations, models, and methods to systematically investigate the structural ordering, force-network topology, and non-equilibrium kinetics of binary d-SALR colloidal assemblies across a spectrum of colloid size ratios. We find that the dynamic persistence of the assembled microsphere clusters and the prevalence of microsphere clusters at high magnetic strengths vary nonmonotonically as function of particle size ratio at different number ratios. To understand how these trends relate to the formation of structural motifs characteristic of alloy phases favored in the binary hard-disk phase diagram, we analyze BD configurations via the classification schemes of Fayen et al.\cite{fayen2020infinite} and Likos and Henley\cite{likos1993complex}, integrated with the analytical pattern detection framework of Lotito and Zambelli\cite{lotito2018pattern, lotito2019journey, lotito2020pattern}. We further characterize the microsphere clusters via the ORC of force-weighted graphs constructed from the simulated configurations.\cite{yeh2026dynamic}. Results from our simulations show a strong correlation between the dynamic persistence of clusters and their propensity to exhibit specific crystalline structural motifs, providing insight into the microscopic factors that govern the stability of d-SALR assemblies. We also observe a strong anti-correlation between cluster lifetimes and ORC, highlighting that force-network topology provides a symmetry-agnostic predictor of nonequilibrium cluster dynamics.

\section{Simulation Methods}

\subsection{\label{sec:model}Model}

\begin{figure}[!htbp]
    \centering
    \includegraphics[width=.8\linewidth]{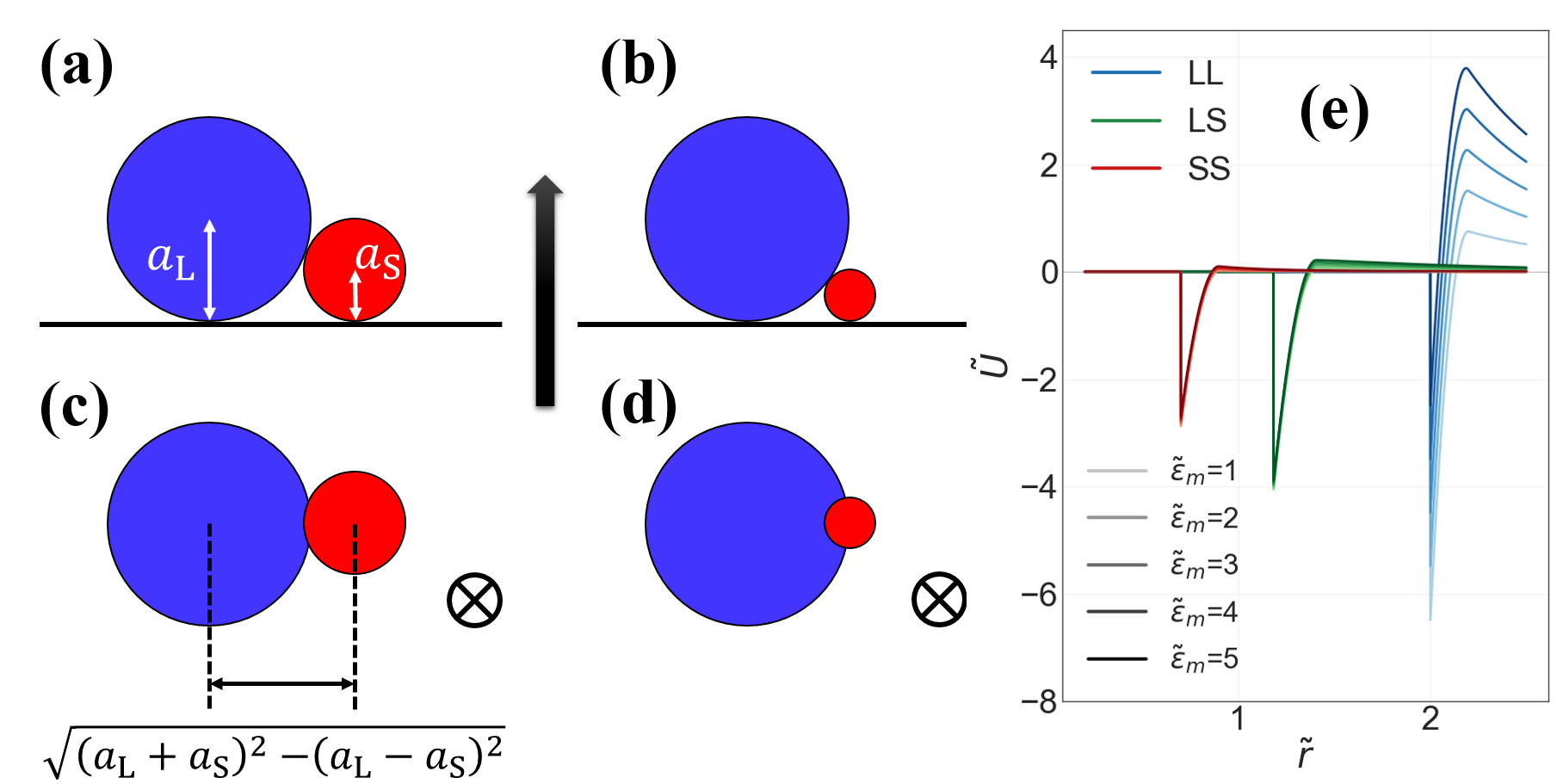}
    \caption{{\bf{Size asymmetry strongly influences quasi-two-dimensional monolayer interactions.}} Side view of large (blue) and small (red) particles on a substrate with size ratios (a) $\tilde a_\text{S}=a_\text{S}/a_\text{L}=0.5$ and (b) $\tilde a_\text{S}=0.25$, respectively. The arrow shows the direction of the applied magnetic field. Bottom view of particles on a substrate with size ratios (c) $\tilde a_\text{S}=0.5$ and (d) $\tilde a_\text{S}=0.25$, respectively. The distance shown $r$ is the interparticle separation projected onto the plane. (e) Pair potential $\tilde{U}$ for LL, LS, and SS microsphere interactions (see Equations~\ref{eq:pair_potent}--\ref{eq:hs_cuts}) as a function plane-projected separation $\tilde{r}=r/a_\text{L}$ for $\tilde a_\text{S}=0.35$ and  $\tilde{\varepsilon}_\text{d}=7.5$ at $\tilde{\varepsilon}_\text{m}=\{1,2,3,4,5\}$, from bright to dark, highlighting how the strength of the magnetic field asymmetrically modifies the interactions.}
    \label{fig:geometry}
\end{figure}

We consider a quasi-two-dimensional binary monolayer of large (L) and small (S) microspheres interacting through a d-SALR potential,\cite{yeh2025colloidal,yeh2026dynamic} in which the magnetic nanoparticles of the surrounding ferrofluid are treated implicitly. In dimensionless form, the pair potential between species \(i,j\in\{{\text{L,S}}\}\) can be expressed
\begin{equation}
\widetilde{U}_{ij}(\tilde r)=
\widetilde{U}_{ij}^{\mathrm{HS}}(\tilde r)+
\widetilde{U}_{ij}^{\mathrm D}(\tilde r)+
\widetilde{U}_{ij}^{\mathrm M}(\tilde r)
\label{eq:pair_potent}
\end{equation}
where \(\widetilde{U}^{\mathrm{HS}}\), \(\widetilde{U}^{\mathrm D}\), and \(\widetilde{U}^{\mathrm M}\) denote contributions from the hard-sphere repulsion, depletion attraction, and magnetic-field-induced dipolar interaction, respectively.  
Particle-center separations projected onto the plane of the substrate are normalized as \(\tilde r=r/a_{\text{L}}\), where \(a_{\text{L}}\) is the radius of the large microspheres, and the size asymmetry or ratio is quantified by the dimensionless radius of the small microspheres, $\tilde{a}_\text{S}=a_\text{S}/a_\text{L}$ (see Figure~\ref{fig:geometry}). Potential energies are normalized by the thermal energy scale \(k_{\mathrm B}T\), where \(k_{\mathrm B}\) is the Boltzmann constant and \(T\) is temperature.  


\subsubsection{Depletion Interaction}

Magnetic nanoparticles in the solvent medium are treated as ideal depletants in the Asakura--Oosawa--Vrij model\cite{Lekkerkerker2024,AOModel1,AOModel2,Vrij}, generating effective short-range attractions between colloids.  
At fixed depletant size \(\tilde a_\text{d}\), the depletion attraction strength between colloids scales with \(\tilde\varepsilon_\text{d}\), a parameter that is experimentally controlled by depletant concentration; the LL, SS, and LS depletion attractions have the following forms,\cite{yeh2026dynamic}
\begin{equation}
\widetilde{U}_{\mathrm{LL}}^{\mathrm D}(\tilde r)=
-\frac{\tilde\varepsilon_{\mathrm d}(\tilde a_{\mathrm d}+1)^3}
{\tilde a_{\mathrm d}^2(1.5+\tilde a_{\mathrm d})}
\left[
1-\frac{3\tilde r}{4(\tilde a_{\mathrm d}+1)}
+\frac{1}{16}\left(\frac{\tilde r}{\tilde a_{\mathrm d}+1}\right)^3
\right]
\label{eq:uLLD_new}
\end{equation}

\begin{equation}
\widetilde{U}_{\mathrm{SS}}^{\mathrm D}(\tilde r)=
-\frac{\tilde\varepsilon_{\mathrm d}(\tilde a_{\mathrm d}+\tilde a_\text{S})^3}
{\tilde a_{\mathrm d}^2(1.5+\tilde a_{\mathrm d})}
\left[
1-\frac{3\tilde r}{4(\tilde a_{\mathrm d}+\tilde a_\text{S})}
+\frac{1}{16}\left(\frac{\tilde r}{\tilde a_{\mathrm d}+\tilde a_\text{S}}\right)^3
\right]
\label{eq:uSSD_new}
\end{equation}

\begin{multline}
\widetilde{U}_{\mathrm{LS}}^{\mathrm D}(\tilde r)=
-\frac{\tilde\varepsilon_{\mathrm d}(\tilde a_{\mathrm d}+\tilde a_{\text{LS}})^3}
{\tilde a_{\mathrm d}^2(1.5+\tilde a_{\mathrm d})}
\Bigg[
1-\frac{3\sqrt{\tilde r^2+\Delta_{\text{LS}}^2}}{4(\tilde a_{\mathrm d}+\tilde a_{\text{LS}})}
+\frac{1}{16}\left(\frac{\sqrt{\tilde r^2+\Delta_{\text{LS}}^2}}{\tilde a_{\mathrm d}+\tilde a_{\text{LS}}}\right)^3 \\
-\frac{3}{16}\frac{\Delta_{\text{LS}}^2}{(\tilde a_{\mathrm d}+\tilde a_{\text{LS}})^3}
\frac{\left(\sqrt{\tilde r^2+\Delta_{\text{LS}}^2}-2(\tilde a_{\mathrm d}+\tilde a_{\text{LS}})\right)^2}
{\sqrt{\tilde r^2+\Delta_{\text{LS}}^2}}
\Bigg]
\label{eq:uLSD_new}
\end{multline}
Here, $\tilde a_{\text{LS}}=(1+\tilde{a}_{\text{S}})/2$, and $\Delta_{\text{LS}}=1-\tilde{a}_{\text{S}}$. The range of the depletion interactions is set by the depletant diameter, and thus the depletion potentials are truncated at the following separations,

\begin{align}
\tilde r_{\mathrm{LL}}^{\mathrm{cut,D}}&=2(1+\tilde a_{\mathrm d}), &
\tilde r_{\mathrm{SS}}^{\mathrm{cut,D}}&=2(\tilde a_\text{S}+\tilde a_{\mathrm d}), &
\tilde r_{\mathrm{LS}}^{\mathrm{cut,D}}&=\sqrt{(1+\tilde a_\text{S}+2\tilde a_{\mathrm d})^2-\Delta_{\text{LS}}^2}
\end{align}

\subsubsection{Magnetic Field Induced Dipolar Interactions}

Under an external field normal to the monolayer, nonmagnetic particles in a ferrofluid acquire induced dipole moments\cite{skjeltorp1983one} and interact via dipolar pair forces\cite{yeh2025colloidal,al2022dual,gauri2023magnetic}.
The resulting LL and SS terms are purely repulsive in-plane and decay as \(\tilde r^{-3}\).  
\begin{equation}
\widetilde{U}_{\mathrm{LL}}^{\mathrm M}(\tilde r)=
\frac{8\tilde\varepsilon_{\mathrm m}}{\tilde r^3}
\label{eq:uLLM_new}
\end{equation}
\begin{equation}
\widetilde{U}_{\mathrm{SS}}^{\mathrm M}(\tilde r)=
\frac{8\tilde\varepsilon_{\mathrm m}\tilde a_\text{S}^6}{\tilde r^3}
\label{eq:uSSM_new}
\end{equation}
For LS pairs, the geometric offset \(\Delta_{\text{LS}}\) modifies both the effective separation and angular factor, leading to interactions with repulsions that decrease with decreasing \(\tilde a_\text{S}\), becoming attractive for \(\tilde a_\text{S} \lesssim 0.268\):
\begin{equation}
\widetilde{U}_{\mathrm{LS}}^{\mathrm M}(\tilde r)=
\frac{8\tilde\varepsilon_{\mathrm m}\tilde a_\text{S}^3}{\left(\sqrt{\tilde r^2+\Delta_{\text{LS}}^2}\right)^3}
\left[
1-3\left(\frac{\Delta_{\text{LS}}}{\sqrt{\tilde r^2+\Delta_{\text{LS}}^2}}\right)^2
\right]
\label{eq:uLSM_new}
\end{equation}
Here \(\tilde\varepsilon_\text{m}\) controls the magnetic interaction strength.  
Because dipole moments scale with particle volumes, SS and LS prefactors are reduced in magnitude relative to LL by factors of \(\tilde a_\text{S}^{-6}\) and \(\tilde a_\text{S}^{-3}\), respectively. 

\subsubsection{Hard-Sphere Interactions}
The microsphere cores contribute hard-sphere-like exclusion interactions, which we treat in BD simulations with time step \(\Delta\tilde t\) using the Heyes-Melrose approach described previously.\cite{heyes1993brownian,yeh2025colloidal,yeh2026dynamic} 
Hard-sphere repulsions for LL and SS pairs are expressed as
\begin{equation}
\widetilde{U}_{\mathrm{LL}}^{\mathrm{HS}}(\tilde r)=
\frac{1}{4\Delta\tilde t}(2-\tilde r)^2,\qquad
\widetilde{U}_{\mathrm{SS}}^{\mathrm{HS}}(\tilde r)=
\frac{1}{4\Delta\tilde t}(2\tilde a_\text{S}-\tilde r)^2
\label{eq:uHS_new_1}
\end{equation} 
The simulation time step $\Delta t = 10^{-4}\tau$ is specified in terms of the diffusion time scale $\tau=\gamma a_\text{L}^2/(k_{\text{B}} T)$, where $\gamma$ is the friction coefficient, and $\tilde{t}=t/\tau$. Each trajectory is run for $2000\tau$, with configurations saved every $4\tau$.
The LS term is nonadditive due to the geometric offset due to the difference in microsphere sedimentation heights,
\begin{equation}
\widetilde{U}_{\mathrm{LS}}^{\mathrm{HS}}(\tilde r)=
\frac{1}{4\Delta\tilde t}
\left[(1+\tilde a_\text{S})-\sqrt{\tilde r^2+\Delta_{\text{LS}}^2}\right]^2
\label{eq:uHS_new_2}
\end{equation}

The hard-sphere interaction cutoffs are given by
\begin{align}
\tilde r_{\mathrm{LL}}^{\mathrm{cut,HS}}&=2, &
\tilde r_{\mathrm{SS}}^{\mathrm{cut,HS}}&=2\tilde a_\text{S}, &
\tilde r_{\mathrm{LS}}^{\mathrm{cut,HS}}&=\sqrt{(1+\tilde a_\text{S})^2-\Delta_{\text{LS}}^2}
\label{eq:hs_cuts}
\end{align}

\begin{figure}[!htbp]
    \centering
    \includegraphics[width=1\linewidth]{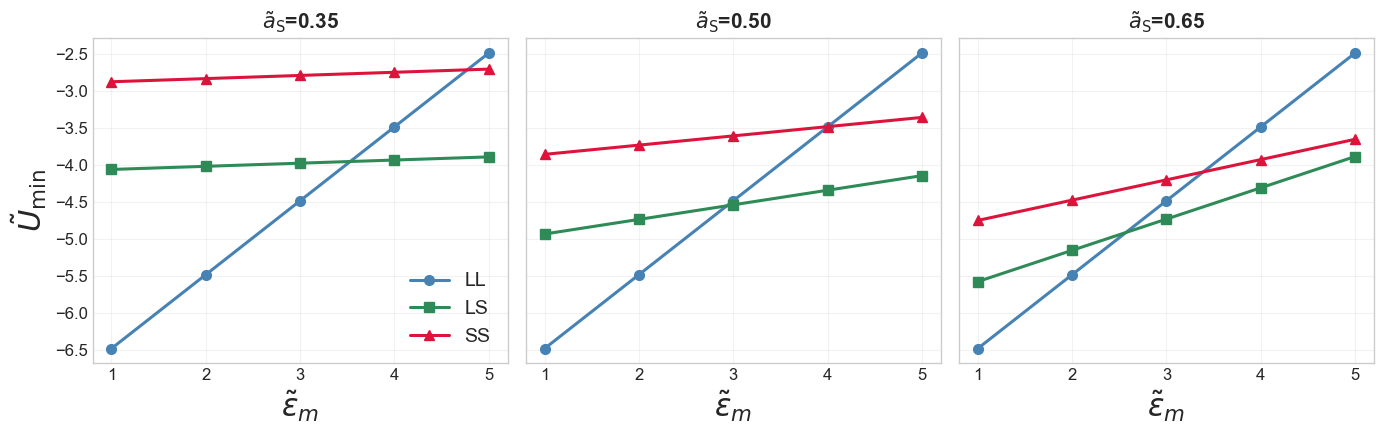}
    \caption{{\bf{Magnetic field strength and microsphere size ratio determine which pair interactions are most attractive.}} 
    Attractive well depth $\widetilde{U}_{\min}$ for LL, LS, and SS interactions between microspheres with size ratios $\tilde a_\text{S}=0.35, 0.50$ and $0.65$ (see Equations~\ref{eq:pair_potent}--\ref{eq:hs_cuts}) as a function of $\tilde{\varepsilon}_\text{m}$ at fixed $\tilde{\varepsilon}_\text{d}=7.5$.}
    \label{fig:umin}
\end{figure}

In summary, \(\tilde\varepsilon_\text{d}\) and \(\tilde\varepsilon_\text{m}\) control the competition between short-range attraction and long-range dipolar repulsion, while \(\tilde a_\text{S}\) and \(\Delta_{\text{LS}}\) establish weights of the dipolar and depletion interactions, as well packing efficiency and geometric nonadditivity.  
These interaction parameters define the binary d-SALR interaction landscape across all size ratios. Figure~\ref{fig:umin} shows that LL pairs exhibit the strongest attractions at low magnetic fields, whereas SS pairs exhibit the weakest. At high fields, this trend reverses, with LS pairs exhibiting the strongest attractions and LL pairs the weakest. As explored below, this crossover produces strong field-dependent trends in structure and dynamics.

\subsection{Brownian Dynamics Simulations}
BD simulations implemented in HOOMD-blue (v2.9.6)\cite{anderson2020hoomd} were used to generate trajectories of microspheres interacting via the model pair potentials given by Equations~\ref{eq:pair_potent}--\ref{eq:hs_cuts}. We simulated $N = 10^4$ colloids in a square, periodically replicated simulation cell of side length \( L\) determined by the desired plane-projected area fraction \(\phi=N \pi(a/L)^2[1-x_{\text{S}}(1-\tilde{a}_{\text{S}}^2)]\), where $x_{\text{S}}=N_{\text{S}}/N$ is the fraction of small microspheres. Other details of the BD simulations are identical to those reported previously.\cite{yeh2025colloidal}

To investigate the effect of particle-size asymmetry on the structures formed, we performed BD simulations at \(\phi=0.25\) considering seven different microsphere size ratios, \(\tilde a_\text{S}=0.25,\ 0.35,\ 0.45,\ 0.50,\ 0.55,\ 0.65,\) and \(0.85\). The ratio \(\tilde a_\text{S}=0.50\) corresponds to that studied in our previous work.\cite{yeh2025colloidal} For each radius ratio studied here, we examined a series of component mixing ratios \(N_\text{L}:N_{\text{S}}\): \(3:1,\ 1:1,\) and \(1:3\), equivalent to \(x_{\text{S}}=0.25,\ 0.5,\) and \(0.75\), denoted here as L3S1, L2S2, and L1S3, respectively. We set the depletion interaction strength to $\tilde{\varepsilon}_\text{d}=7.5$ for all simulations, a value which we have previously shown leads to cluster-size distributions in this model that are consistent with experiments of a d-SALR fluid with $\tilde a_\text{S}=0.5$ and the magnetic-field-induced dipolar interactions considered here, $\tilde{\varepsilon}_\text{m} \in \{1,2,3,4,5\}$.\cite{yeh2026dynamic}

\subsection{Structural Motifs}

\begin{figure}[!htbp]
    \centering
    \includegraphics[width=1\linewidth]{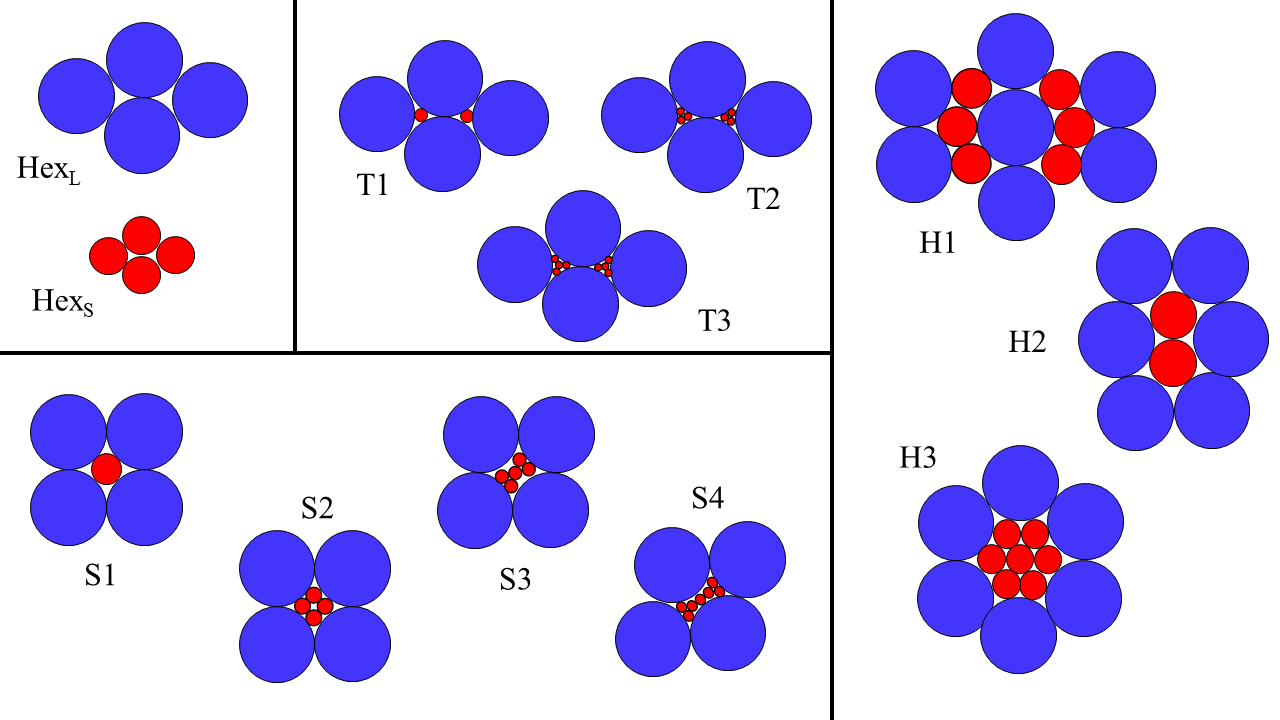}
    \caption{{\bf{Two-dimensional structures favored in binary sphere mixtures.}} Local packing arrangements preferred for different particle radius ratios (additive interaction case shown). Adapted from ref.\citenum{fayen2020infinite}.}
    \label{fig:nonaddself}
\end{figure}

To identify local structures favored by the binary d-SALR mixtures studied here, we developed a classifier that detects motifs based on their defining contact topology, rather than by neighbor-count or bond-orientational-order heuristics \cite{likos1993complex,fayen2020infinite}. Particle configurations were used to construct contact networks, considering particle pairs separated by distances less than $\tilde r_{\mathrm{ij}}^{\mathrm{cut,HS}}+0.075$, where \(i,j\in\{L,S\}\).
These networks were used to create bond graphs, $G_\text{LL}$ and $G_\text{SS}$, and a bipartite L--S contact list, from which every candidate motif (see Figure~\ref{fig:nonaddself}) was described as a closed ring or compact cluster. Square-family (S1, S2) and rhombic (T1) motifs were identified by exhaustively enumerating 4-membered cycles in $G_\text{LL}$ and classifying each cycle's interior angles and opposite-side length ratios as rectangular or parallelogrammatic. The number and connectivity of S particles enclosed within the ring (tested via point-in-polygon membership) then distinguished S1 (one central S bonded to all four ring vertices), S2 (four S particles each bridging two ring vertices and two neighboring S), and T1 (two enclosed S particles within a non-rectangular parallelogram). Triangular motifs (T2, T3) were instead identified from connected components of the S--S contact graph. Compact three- or four-particle S clusters that were fully enclosed by exactly three mutually bonded L particles were classified as T2 or T3, respectively. Honeycomb-derived motifs (H1--H3) were detected as closed six-membered L rings enclosing a fixed number of S particles with a prescribed bonding pattern. To exclude rings that geometrically degenerate into a four-sided shape, each candidate hexagonal ring was additionally required to have all six vertices lying on its convex hull. Finally, residual L or S particles exhibiting complete six-fold same-species coordination with negligible cross-species contact were grouped by connectedness into bulk hexagonal/triangular lattice domains ($\mathrm{HEX_{\text{L}}}$, $\mathrm{HEX_{\text{S}}}$), representing the pure single-species crystalline background. Because a given particle may satisfy the coordination requirements of more than one candidate motif, structures were assigned greedily in order of specificity (S/T motifs, then triangular clusters, then honeycomb motifs, then the bulk hexagonal background), and once a particle was claimed by a confirmed structure it was excluded from all subsequent searches, guaranteeing a unique, non-overlapping structural assignment for every particle in the system. All ring and cluster geometries were evaluated in a locally unwrapped reference frame obtained via minimum-image displacement, ensuring that motifs spanning periodic boundaries were correctly resolved. This procedure yielded, for each configuration, an unambiguous census of the ten crystalline phase populations (S1, S2, T1, T2, T3, H1, H2, H3, $\mathrm{HEX_{\text{L}}}$, $\mathrm{HEX_{\text{S}}}$), which we accumulated over the last ten equilibrated trajectory frames of each state point to obtain statistically representative phase-count maps across the size-ratio ($\tilde a_\text{S}$) and composition ($x_\text{S}$) parameter space.

As a point of comparison, we also computed local bond-orientational order parameters to quantify the degree of crystalline ordering and distinguish between different emergent symmetries. Specifically, we consider both four-fold ($\psi_4$) and six-fold ($\psi_6$) bond-orientational order parameters, which characterize square-like and hexagonal ordering, respectively.\cite{glaser1992melting} The global order parameter is defined as
\begin{equation}
\psi_n = \frac{1}{N} \sum_{j=1}^{N} |\psi_{n,j}|, \qquad n=4,6,
\end{equation}
where $\psi_{n,j}$ represents the local bond-orientational order of particle $j$, defined as
\begin{equation}\label{eq:psin}
\psi_{n,j} = \frac{1}{N_j} \sum_{k=1}^{N_j} \exp\left(i n \theta_{jk}\right), \qquad n=4,6.
\end{equation}
Here, $N$ is the number of particles, $N_j$ is the number of neighbors of particle $j$, and $\theta_{jk}$ is the angle between the bond connecting particles $j$ and $k$ and a fixed reference axis. The magnitude of $\psi_{n,j}$ measures the extent of local $n$-fold orientational order, while the global parameter $\psi_n$ characterizes its persistence throughout the structure.

\subsection{Force-Weighted Interaction Graphs and Ollivier--Ricci Curvature}

Graph-theoretic descriptors, including those with network edges that are weighted to encode relevant physics, provide a powerful means of characterizing topological order in particle-based materials that can be otherwise difficult to capture using local or symmetry-based order parameters  \cite{jiang2020emergence,vecchio2021structural,vecchio2022spanning,papadopoulos2018network,hallstrom2026decoding,yeh2026dynamic}. 
To understand how structure affects heat transfer in granular materials, graphs with edges weighted by local heat-transfer coefficients have been productively employed.\cite{smart2007effects}. Following this approach, each monolayer configuration was represented here as a force-weighted graph in which vertices are individual particles (L or S) and edges connect pairs experiencing a non-negligible interaction force, combining depletion, magnetic, and steric/contact contributions; edge weights are set by the instantaneous pairwise force magnitude, and an inverse-force edge length is used to give the graph a metric structure, as in our previous work \cite{yeh2026dynamic}.

On these weighted graphs, we computed the Ollivier--Ricci curvature (ORC) of each edge, $\kappa(u,v)$, following the same construction (force-weighted neighbor measures and an optimal-transport/Wasserstein comparison of those measures) used previously \cite{ollivier2009ricci,sia2019ollivier,eidi2020ollivier,jamadgni2026graph,gosztolai2021unfolding}; we do not repeat the full definitions here. Physically, positive-curvature edges mark force pathways embedded in cooperative, overlapping neighborhoods capable of sharing and redistributing stress, while negative curvature identifies bridge-like edges linking mechanically distinct regions, and near-zero curvature reflects a locally homogeneous, lattice-like force network \cite{jamadgni2026graph}. As before, curvature values are separately quantified by contact type (LL, LS, SS) to distinguish disparate contributions of each bond type to local reinforcement versus long-range stress transmission across the various combinations of depletion- and dipolar-interaction strengths explored in this work. They are then averaged over edges to form nodal ORC for each microsphere.

\subsection{Cluster Persistence}

Microsphere cluster lifetimes are quantified via the fractional similarity of associates in a cluster (FSAC) autocorrelation function, $\Phi(\tilde{t})$, which measures the fraction of a particle's initial cluster associates that remain shared at a later time $\tilde{t}$ \cite{jadrich2015equilibrium,yeh2026dynamic}.
This quantity can be expressed as
\begin{equation}
    \Phi(\tilde{t})=\frac{1}{N}\sum_{i=1}^{N} \left[\frac{\Phi_{i,\text{shared}}(\tilde{t})}{\Phi_{i,\text{total}}(\tilde{t})}\right]
\label{eq:fsac}
\end{equation}
where $\Phi_{i,\mathrm{shared}}(\tilde{t})$ is the number of associates common to the cluster of particle $i$ at both the initial time and time $\tilde{t}$, while $\Phi_{i,\mathrm{total}}(\tilde{t})$ is the total number of unique associates across the two time points. By this measure, $\Phi(\tilde{t})=1$ corresponds to perfectly persistent cluster membership, and $\Phi(\tilde{t})\to0$ indicates complete membership turnover. 
From each $\Phi(\tilde{t})$, we extracted a single characteristic relaxation time, $\tilde \tau_{\mathrm{char}}$, where $\Phi(\tilde{\tau}_{\mathrm{char}})=e^{-1}$, which we used throughout as the scalar measure of cluster persistence. 

\subsection{Statistical Correlation of Structure and Dynamics}
To quantify how ORC and the local structural motif populations relate to dynamic cluster persistence, we computed the Spearman rank correlation coefficient $\rho$ between each descriptor and $\tilde{\tau}_{\mathrm{char}}$. Each coefficient was computed pairwise using only available paired observations for the descriptor,
together with the associated two-sided $p$-value testing against the null hypothesis (no statistical correlation).  Descriptors for which fewer than three valid observations were available, or whose values were constant across the dataset, were excluded from the analysis. The sign of $\rho$ indicates the direction of the association (positive: descriptor increases with $\tilde{\tau}_{\mathrm{char}}$; negative: descriptor decreases with $\tilde{\tau}_{\mathrm{char}}$), while its magnitude reflects the strength of the monotonic trend, allowing descriptors to be ranked by their univariate relevance to the FSAC response.

\section{Simulation Results}
\subsection{Simulated Configurations}

\begin{figure}
    \centering
    \includegraphics[width=1\linewidth]{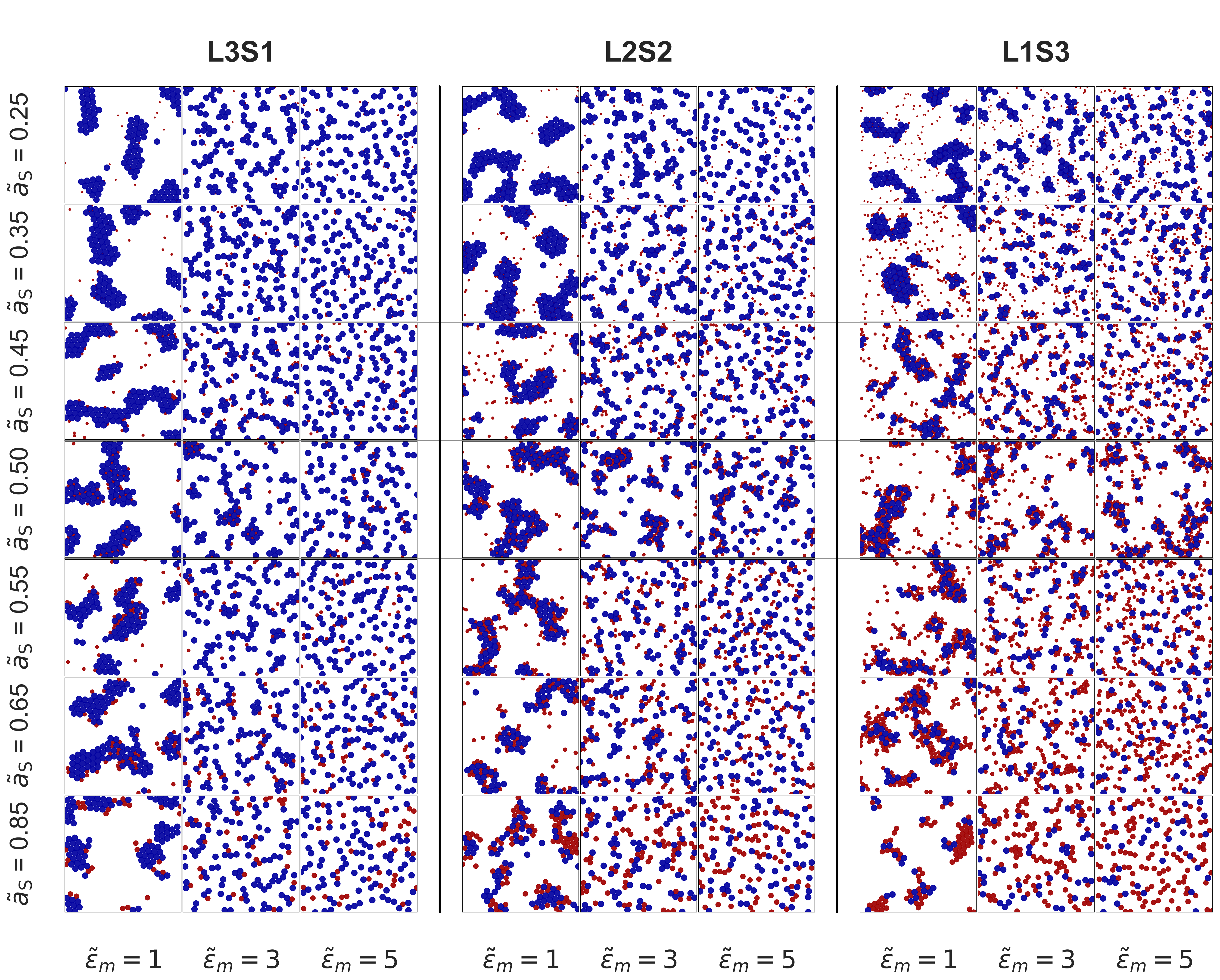}
    \caption{{\bf{Binary cluster morphology is tunable by magnetic field strength, microsphere size, and microsphere mixing ratio.}} Zoomed-in configurations from BD simulations grouped into three main columns by microsphere mixing ratio (L3S1, L2S2, L1S3) each with subcolumns corresponding to \(\tilde{\varepsilon}_\text{m}=1,3,5\). Rows correspond to increasing small-particle radius ratio \(\tilde a_\text{S}\), as labeled on the left. Each panel shows large particles (blue, radius \(\tilde{a}_{\text{L}}=1\)) and small particles (red, radius \(\tilde a_\text{S}\)). This layout enables direct visual comparison of morphology across size asymmetry and magnetic coupling at fixed depletion interaction strength, $\tilde{\varepsilon}_\text{d}=7.5$.}
    \label{fig:bd_snapshots_custom_grid}
\end{figure}

Microsphere configurations obtained from BD simulations at different dipolar interaction strengths $\tilde{\varepsilon}_\text{m}$, particle size ratios $\tilde a_\text{S}$, and mixing number ratios (L3S1, L2S2, and L1S3) illustrate how these independent variables control structure formation in d-SALR colloids (see Figure \ref{fig:bd_snapshots_custom_grid}). 
The primary effect of increasing $\tilde{\varepsilon}_\text{m}$ is enhancing the dipolar repulsions between large microspheres, balancing and ultimately overcoming their depletion attractions to reduce microsphere cluster size. Increasing $\tilde{\varepsilon}_\text{m}$ also transforms how the clusters organize due to its asymmetric, size-ratio-dependent effect on net LL, LS, and SS interactions. Since the magnitude of induced magentic dipoles scales with particle volume, the LS and SS pair interactions have weaker $\tilde{\varepsilon}_\text{m}$ dependencies compared to those between large microspheres (see Figure~\ref{fig:umin}). One consequence, demonstrated recently for the case of $\tilde a_\text{S}=0.5$\cite{yeh2026dynamic} and also apparent across the size ratios considered here, is that clustering persists to higher $\tilde{\varepsilon}_\text{m}$ in microsphere mixtures of large and small particles, dominated by cross-species attractions where small microspheres link together large particles. This effect, which trivially disappears when particle radius ratios approach unity, also becomes less important as $\tilde a_\text{S}$ approaches zero, where species segregation (selective crystallization) is instead favored (see Figure~\ref{fig:bd_snapshots_custom_grid}), as has been reported for three-dimensional binary colloidal suspensions interacting via depletion attractions.\cite{yang2026fractional} The striking differences between the dynamic structures formed at various intermediate size ratios between these limits is one of the main results of this study, highlighting the influence of microsphere radius ratio as a control parameter for d-SALR colloids.

\subsection{Structural Characterization}

\begin{figure}[!htbp]
    \centering
    \includegraphics[width=1\linewidth]{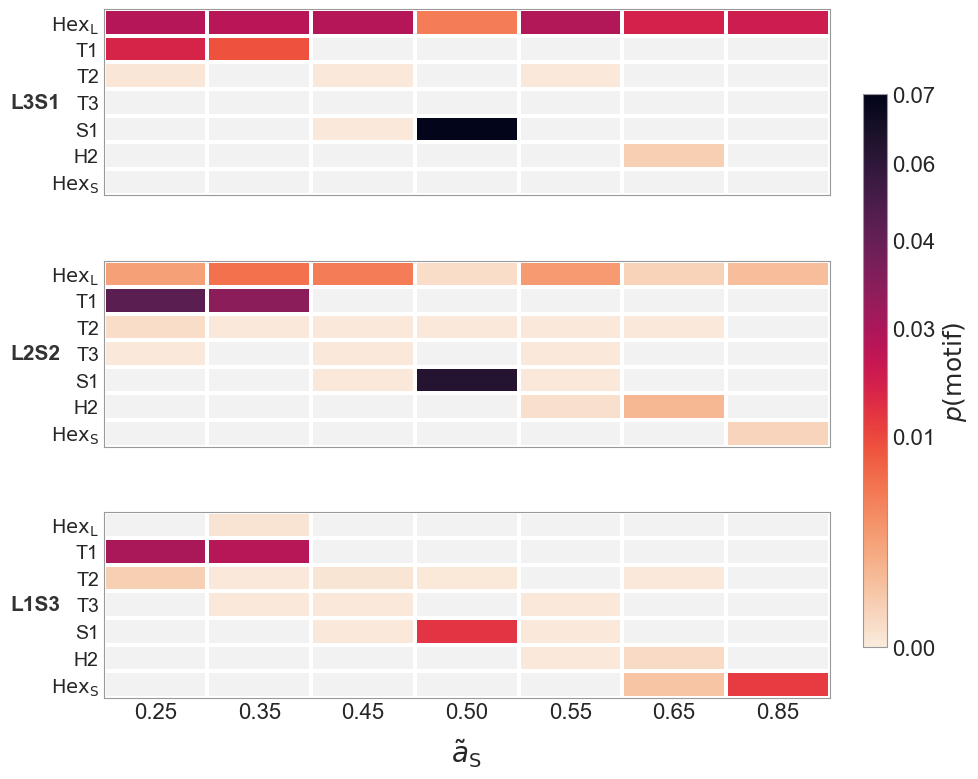}
    \caption{{\bf{Size ratio and mixture composition determine the prevalence of crystal motifs in simulated binary d-SALR  microsphere clusters.}} Probability that a randomly chosen particle is associated with a local packing motif from Figure~\ref{fig:nonaddself}, as a function of microsphere size ratio $\tilde a_\text{S} = a_{\text{S}}/a_{\text{L}}$ and binary mixture composition (mixing ratio).
    The three stacked panels correspond to large-to-small number ratios $3\!:\!1$ (L3S1, top), $1\!:\!1$ (L2S2, middle), and $1\!:\!3$ (L1S3, bottom). Rows index the ten identified motifs
        [$\mathrm{Hex_{\text{L}}}$, $\mathrm{T1}$, $\mathrm{T2}$, $\mathrm{T3}$,
        $\mathrm{S1}$, $\mathrm{S2}$, $\mathrm{H1}$, $\mathrm{H2}$, $\mathrm{H3}$,
        $\mathrm{Hex_{\text{S}}}$], and columns indicate the radius ratio $\tilde a_\text{S}$.
        Color indicates $p(\mathrm{motif})$, the probability is a given particle is in motif $i$,
        averaged over the last ten equilibrium BD configurations at $\tilde{\varepsilon}_\text{m} = 1$ and $\tilde{\varepsilon}_\text{d}=7.5$.
        By construction, $\sum_i p(\mathrm{motif}) = 1 - p(\mathrm{no\ motif})$,
        so uncolored probability at a given $\tilde a_\text{S}$ corresponds to particles
        not classified into any of the ten motifs.
    }
    \label{fig:crystalline}
\end{figure}

To understand how microsphere radius ratio and mixture composition influence the internal organization of clusters formed due to depletion attractions in our model binary d-SALR colloids, we identified the crystalline motifs of Figure~\ref{fig:nonaddself} that were present in BD simulations at $\tilde\varepsilon_\text{m}=1$ (see Figure~\ref{fig:crystalline}). The most striking feature is the preference for S1 structures at $\tilde a_\text{S}=0.5$, where one S particle just fits into the interstitial space of a four-L-particle square ring, a configuration with 4 LL contacts and 4 LS contacts that maximizes their respective depletion attractions. Although S1 is the most common motif across all mixing ratios at $\tilde{a}_\text{S}=0.5$, it does not emerge when the size asymmetry is high ($\tilde{a}_\text{S}\lesssim0.35$) or low ($\tilde{a}_\text{S}\gtrsim0.65$). In the case of the former, the prominent motifs were close-packed hexagonal arrangements of L particles, either with small particles in the interstices (T1) at high $x_{\text{S}}$ or without interstitial particles (Hex$_{\text{L}}$) at low $x_{\text{S}}$. For low asymmetry mixtures $\tilde a_\text{S}\gtrsim0.65$, as expected, close-packed hexagonal arrangements, $\mathrm{HEX_{\text{L}}}$ or $\mathrm{HEX_{\text{S}}}$, are the most common at low or high low $x_{\text{S}}$, respectively. Taken together, these trends indicate that microsphere size ratio primarily selects the optimal symmetries of the host-guest structural families, while mixture composition determines whether a single-component lattice ($\mathrm{HEX_{\text{L}}}$ or $\mathrm{HEX_{\text{S}}}$) (i.e., selective crystallization) occurs. To connect with other symmetry-based structural metrics, we also computed both square ($\psi_4$) and hexatic ($\psi_6$) bond-orientational order parameters for these configurations. While these measures provide less detail than the motif detection approach described above, they reinforce the role of radius ratio in determining prevalent square or hexagonal symmetries, and in particular the strong preference for square ordering at $\tilde a_\text{S}=0.5$, where the S1 motif is most abundant (see Figure S1).

\subsection{Ollivier-Ricci Curvature on Force-Weighted Graphs}
Graphs with colloids represented as nodes and connected by edges weighted by the magnitude of the pairwise force between them were recently used to characterize the interaction topology of quasi-two-dimensional binary d-SALR microspheres with $\tilde a_\text{S}=0.50$.\cite{yeh2026dynamic} Graph-theoretic approaches for describing such complex colloidal networks, which exhibit structures that are neither fully ordered or disordered, are advantageous in that they are not constrained to considering only local structural features nor do they require a priori knowledge of relevant crystalline symmetries.\cite{hallstrom2026decoding} In the case of binary d-SALR colloids with $\tilde a_\text{S}=0.5$, ORC of the network---whether averaged over all edges or locally to compute nodal ORC associated with the large or small particles---was strongly correlated with the particle exchange kinetics of colloidal clusters (i.e., anti-correlated with cluster persistence).\cite{yeh2026dynamic} Here, to understand the effects of size ratio on cluster interaction topology for binary d-SALR mixtures, we compute the nodal ORC as a function of $\tilde a_\text{S}$ for three mixing ratios (L3S1, L2S2, and L1S3), considering $\tilde\varepsilon_\text{m}=1$ and $\tilde{\varepsilon}_\text{d}=7.5$ (see Figures S2-S8).   

\begin{figure}[!htbp]
    \centering
    \includegraphics[width=1\linewidth]{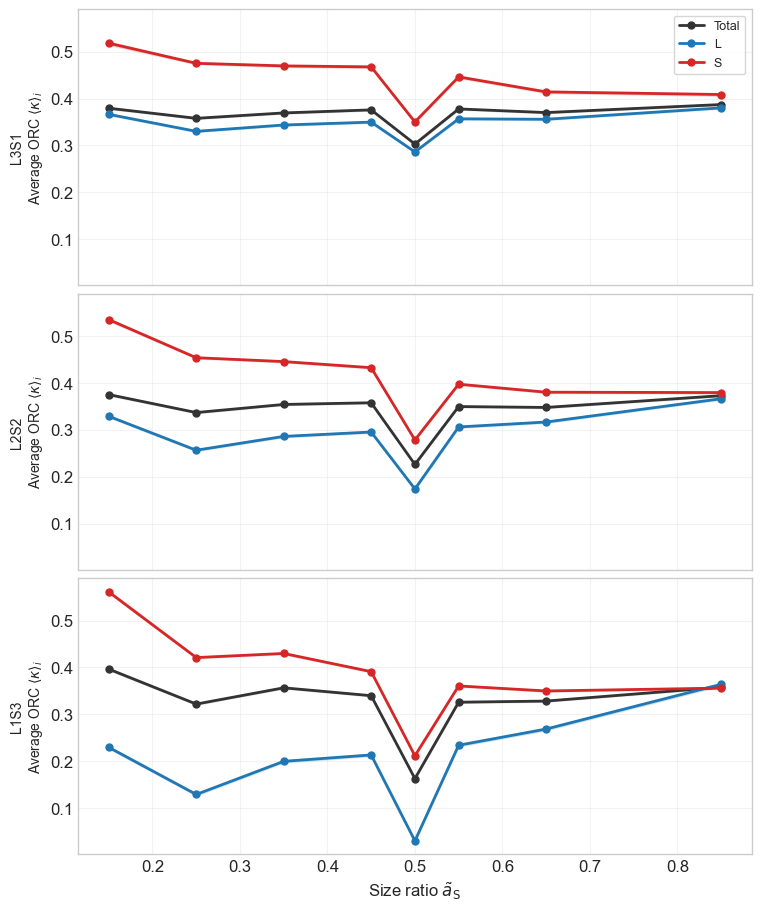}
    \caption{{\bf{Nodal Ollivier--Ricci curvature (ORC) quantifies interaction topological differences in force-magnitude-weighted graphs for simulated binary d-SALR colloids.}} Shown are the nodal ORC $\langle \kappa \rangle_i$, where $i$ indicates an average over either large-particle (L), small-particle (S), or total (L+S) nodes for $\tilde\varepsilon_\text{m}=1$ and $\tilde{\varepsilon}_\text{d}=7.5$ and three compositions: L3S1, L2S2, and L1S3 (top to bottom panels).}
    \label{fig:orc_mean}
\end{figure}

Analysis of the simulated configurations show that nodal ORC is positive $\langle\kappa\rangle\approx0$--$0.60$ across the parameter range explored (Figure~\ref{fig:orc_mean}) as may be expected given the dense network of triangles in the close-packed interior of cluster graphs. Species-resolved mean curvatures $\langle \kappa \rangle_i$, obtained by averaging the nodal ORC over all particles of type $i$, reveal a systematic asymmetry: small particles consistently exhibit more positive average curvature than large particles. As $\tilde a_\text{S}\to0.85$ (the differences between large and small particles begin to vanish), and the large- and small-particle curvatures nearly converge.

The most prominent feature of ORC occurs at $\tilde a_\text{S}=0.50$, where the total, large-, and small-particle averaged curvatures all drop sharply relative to the other size ratios investigated. This feature coincides with the emergence of square ordering of large particles within the clusters with small particles in the interstices (S1). This particular type of square ordering, and the associated drop in ORC for small and large particles, corresponds to the formation of a more evenly distributed force network.  

\subsection{Cluster Persistence}
\begin{figure}[!htbp]
    \centering
    \includegraphics[width=1\linewidth]{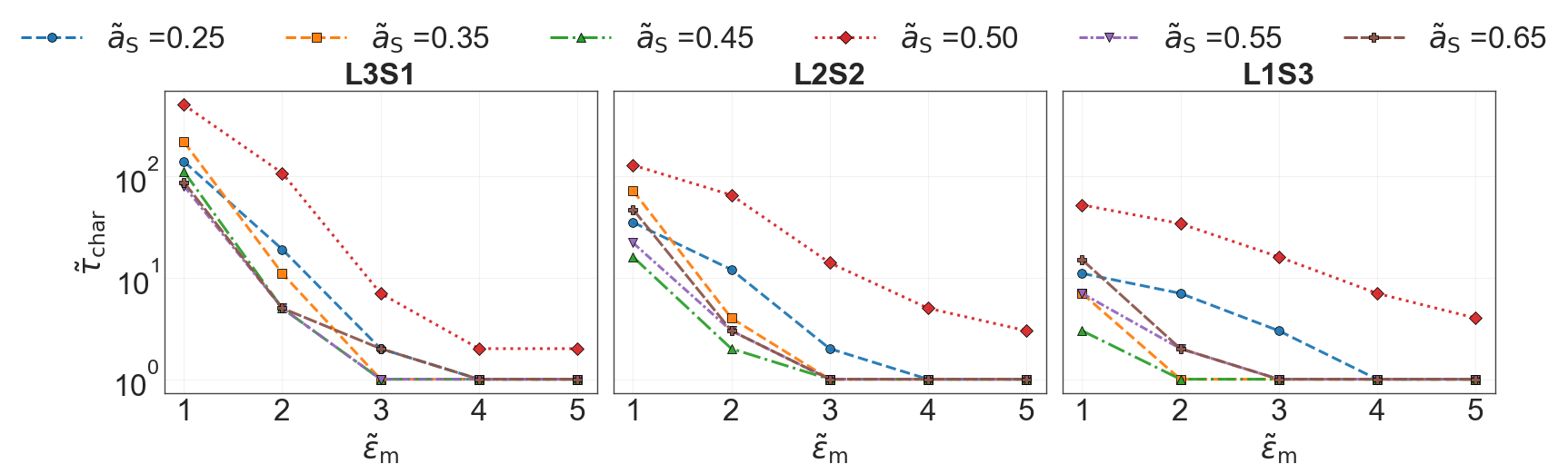}
    \caption{{\bf{Microsphere size ratio and monolayer mixture composition set the magnitude and magnetic-field sensitivity, respectively, of the cluster persistence time $\tilde \tau_{\rm char}$}}. The characteristic time of the monolayer FSAC for simulated binary d-SALR colloids is shown as a function of size ratio $\tilde a_\text{S}$ and dipolar interaction strength $\tilde{\varepsilon}_\text{m}$ for three monolayer mixing ratios: L3S1, L2S2, and L1S3. Within each panel, curves correspond to selected size ratios $\tilde a_\text{S}=0.25,\,0.35,\,0.45,\,0.50,\,0.55,\,0.65$, shown with different color. Data for $\tilde a_\text{S}=0.50$ were reported in a previous publication~\cite{yeh2026dynamic}.}
    \label{fig:fsac}
\end{figure}

To determine how microsphere size ratio modulates cluster persistence, we computed the temporal decay of the fractional similarity of associates in a cluster (FSAC) (see Figures S9-S14) and extracted characteristic times to quantify how cluster exchange kinetics depend on size ratio~$\tilde a_\text{S}$ and dipolar interaction strength $\tilde{\varepsilon}_\text{m}$ for three monolayer compositions (L3S1, L2S2, L1S3) (Figure~\ref{fig:fsac}). For all conditions studied, $\tilde \tau_\text{char}$ decreased monotonically with increasing $\tilde{\varepsilon}_\text{m}$, confirming that field-induced dipolar repulsion promotes cluster turnover regardless of the specific size disparity between species.  The magnitude of the characteristic time, however, depended nonmonotonically on the size ratio $\tilde a_\text{S}$, with the most persistent clusters observed for $\tilde a_\text{S}=0.5$, a size ratio that promotes relatively high S1 order and has force-magnitude-weighted graphs characterized by relatively low ORC. The large-to-small mixing ratio also plays an important role, modifying the sensitivity of $\tilde \tau_\text{char}$ to $\tilde{\varepsilon}_\text{m}$. When large particles are the majority species, they form the most persistent clusters at low $\tilde{\varepsilon}_\text{m}$ because of strong depletion attractions between large microspheres. However, their cluster dynamics are also the most sensitive to changes in $\tilde{\varepsilon}_\text{m}$, due to the large magnitude of the magnetic-field-induced dipolar repulsions between the large particles. For the same reasons, monolayers with a majority of small particles exhibit the least persistent clusters at low $\tilde{\varepsilon}_\text{m}$ and the most persistent clusters at high $\tilde{\varepsilon}_\text{m}$.

\paragraph{Correlation of cluster persistence with crystalline motifs.}
Having characterized how binary crystalline motif propensities varied with size ratio $a_{\text{S}}$ and mixing composition $x_{\text{S}}$ (Figure~\ref{fig:crystalline}), we next examined how they correlated with cluster turnover time as quantified by $\tilde \tau_{\rm char}$ across all state points investigated, using a Spearman rank correlation test (Figure~\ref{fig:placeholder}). A strong monotonic correlation ($|\rho|\gtrsim0.60$) was observed between  $p(\mathrm{Hex_{\text{L}}})$ and $\tilde \tau_{\rm char}$ ($\rho=0.69$), while $p(\mathrm{S_1})$ and $\tilde \tau_{\rm char}$ were only moderately correlated ($\rho=0.50$).  The remaining motif probabilities -- $p(\mathrm{T1})$, $p(\mathrm{T2})$, $p(\mathrm{T3})$, $p(\mathrm{H1})$, $p(\mathrm{Hex_{\text{S}}})$ -- exhibited weak, statistically insignificant correlations with cluster persistence ($|\rho|\le0.16$). We found a strong anticorrelation between cluster persistence and $p(\mathrm{no\ motif})$ ($\rho=-0.79$), indicating a physically intuitive result that increased structural disorder correlates with faster cluster turnover. The magnitude of this correlation is similar to that of the anticorrelation between cluster persistence and dipolar field strength $\tilde\varepsilon_\text{m}$ ($\rho=-0.76$). Due to its strongly nonmonotonic relation to dynamic relaxation, $\tilde a_\text{S}$ has a near-zero Spearman correlation coefficient with $\tilde \tau_{\rm char}$.



\begin{figure}[!htbp]
    \centering
    \includegraphics[width=1\linewidth]{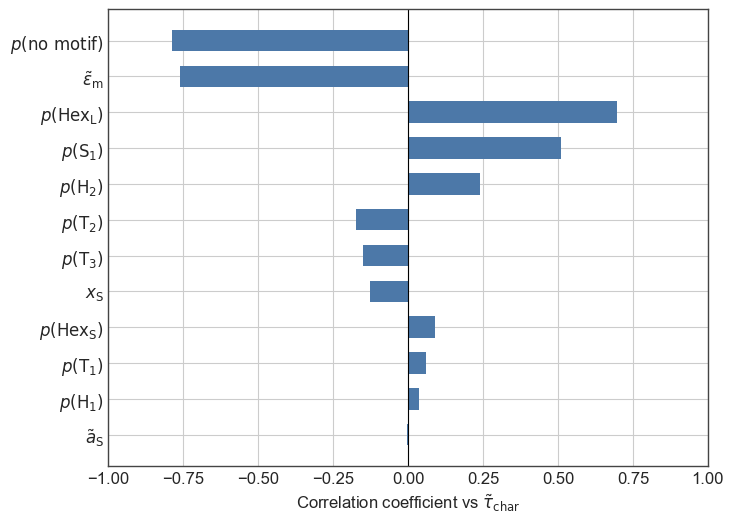}
    \caption{{\bf{Spearman rank correlation coefficients between simulated crystalline-motif probabilities and the characteristic cluster persistence time.}} Coefficients for how other system parameters (small-sphere fraction $x_\text{S}$, dipole interaction strength $\tilde\varepsilon_\text{m}$ and radius ratio $\tilde a_\text{S}$) correlate with $\tilde \tau_{\rm char}$ are provided for comparison.}
    \label{fig:placeholder}
\end{figure}


\paragraph{Correlation of cluster persistence with Ollivier-Ricci curvature.}
Spearman rank correlation coefficients computed between the total and particle-based ORC descriptors and $\tilde \tau_{\mathrm{char}}$ also revealed strong anticorrelations when pooled over all simulated conditions. The mean network curvature $\langle\kappa\rangle$ showed the strongest anticorrelation ($\rho=-0.89$), while the nodal ORCs for large and small particles ($\langle\kappa\rangle_\text{L}$ and $\langle\kappa\rangle_\text{S}$) were also strongly anticorrelated with $\tilde \tau_{\mathrm{char}}$ ($\rho=-0.78$ and $\rho=-0.76$, respectively).  This pattern indicates that modifications to d-SALR colloidal monolayers that increase their local curvature tend to produce states with faster cluster exchange kinetics and less persistent clusters. This correlation is intriguing because it suggests that interaction topological metrics, which do not require information about preferred symmetries, provide information useful for predicting the dynamic properties of complex colloidal assemblies.   
\section{Experimental observations of local packing and cluster formation}
To assess whether the size ratio dependence of cluster propensity observed in the simulations is also evident experimentally, we investigated 2D binary colloidal monolayers immersed in a ferrofluid using a previously reported experimental platform.\cite{al2022dual, gauri2023magnetic} We dispersed non-magnetic microparticles in an aqueous ferrofluid (PBG-200, Ferrotec corp.) containing iron oxide nanoparticles (diameter \ensuremath{\sim}\,12 nm). 
By combining microparticles of different sizes, we investigated how the small-to-large particle radius ratio affects assembly under d-SALR interactions.

We used carboxylated polystyrene latex microspheres of well-defined sizes as model nonmagnetic colloids. The particles were purchased from Magsphere Inc. and Invitrogen and washed with deionized water (electrical resistivity = 18.2 M\ensuremath{\Omega\cdot}cm) to remove any contaminants and excess surfactants. Particle size distributions were determined from optical and electron microscopy images. For particle size measurements, a droplet of the submicrometer particle stock suspension was dried on a piece of silicon wafer. We imaged the particles by scanning electron microscopy (SEM) using a Quanta 3D DualBeam FEG FIB-SEM. The dried specimens were coated with a 5 nm layer of platinum to reduce charging during imaging. SEM images used to quantify particle diameters  were acquired at accelerating voltage of 20 kV. The particles with diameter $>1$ µm, in-situ bright field microscopy (details provided below) was used without drying the dispersion. Both for SEM and optical microscopy, ImageJ was used for obtained the particle sizes. The four particle populations had mean diameters of 0.47 \ensuremath{\pm} 0.02, 1.15 \ensuremath{\pm} 0.09, 1.50 \ensuremath{\pm} 0.08, and 2.21 \ensuremath{\pm} 0.15~\ensuremath{\mu}m (Figure~S15). Values are reported as mean \ensuremath{\pm} standard deviation of the particle diameter, with both quantities obtained from lognormal fits to the measured size distributions.

In a typical assembly experiment, a binary particle mixture ($x_\text{S}=0.5$, L2S2) was dispersed in an aqueous ferrofluid and loaded into a flat borosilicate capillary (VitroCom 5010). The capillary was mounted on a glass slide and sealed at its two ends with hot-melt adhesive. After mounting, the microparticles were allowed to sediment overnight to form a 2D monolayer. The structures formed in the absence of field were imaged using SEM.  To prepare the binary assemblies for SEM, a droplet of the particle mixture dispersed in aqueous ferrofluid was deposited on a piece of silicon wafer. The wafer was placed in a Petri dish containing water to decrease the rate of solvent evaporation and minimize drying-induced flows. Note that drying can generate lateral capillary forces that can rearrange particles which will not be completely eliminated by slow drying. Here the binary particle assembly was imaged in secondary-electron mode of the SEM mode using an Everhart-Thornley detector at 20 kV. For all our experiments, the total 2D projected particle area fraction (defined as the fraction of the analyzed substrate area occupied by the projected footprints of both particles) was kept fixed to $\approx0.4$. 
\begin{figure}[!htbp]
  \centering
  \includegraphics[width=\linewidth]{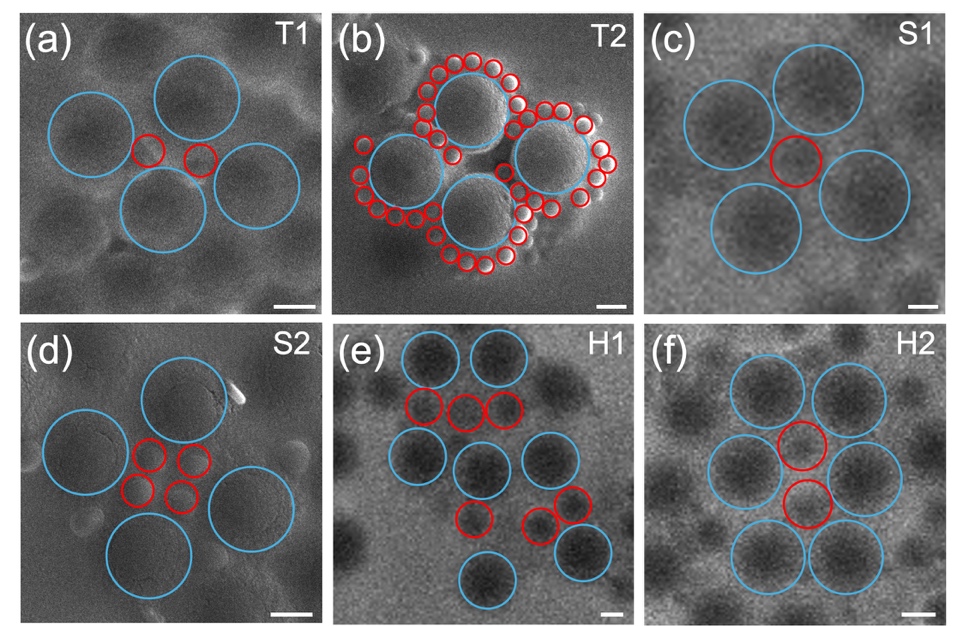}
  \caption{\textbf{Local packing motifs in binary colloidal assemblies.} Representative scanning electron microscopy (SEM) images show (a) T1, (b) T2 (see Figure S16(c)), (c) S1, (d) S2, (e) H1, and (f) H2. Cyan and red outlines mark the large and small microspheres, respectively. The scale bar in each image represents 1~\ensuremath{\mu}m.}
  \label{fig:experimental-packing}
\end{figure}
Representative SEM images (Figure~\ref{fig:experimental-packing}) show local packing arrangements that resemble motifs observed in the simulations (Figure~\ref{fig:nonaddself} and~\ref{fig:crystalline}), including T1, T2, S1, S2, H1, and H2. Note that the motif with small particles closely surrounding triagonally-arranged large particles (Figure~\ref{fig:experimental-packing}(b)) is one of several more complex forms of T2 ordering reported earlier\cite{likos1993complex} (see also Figure S16(c)), but not explicitly shown in Figure~\ref{fig:crystalline}.

\begin{figure}[!htbp]
  \centering
  \includegraphics[width=\linewidth]{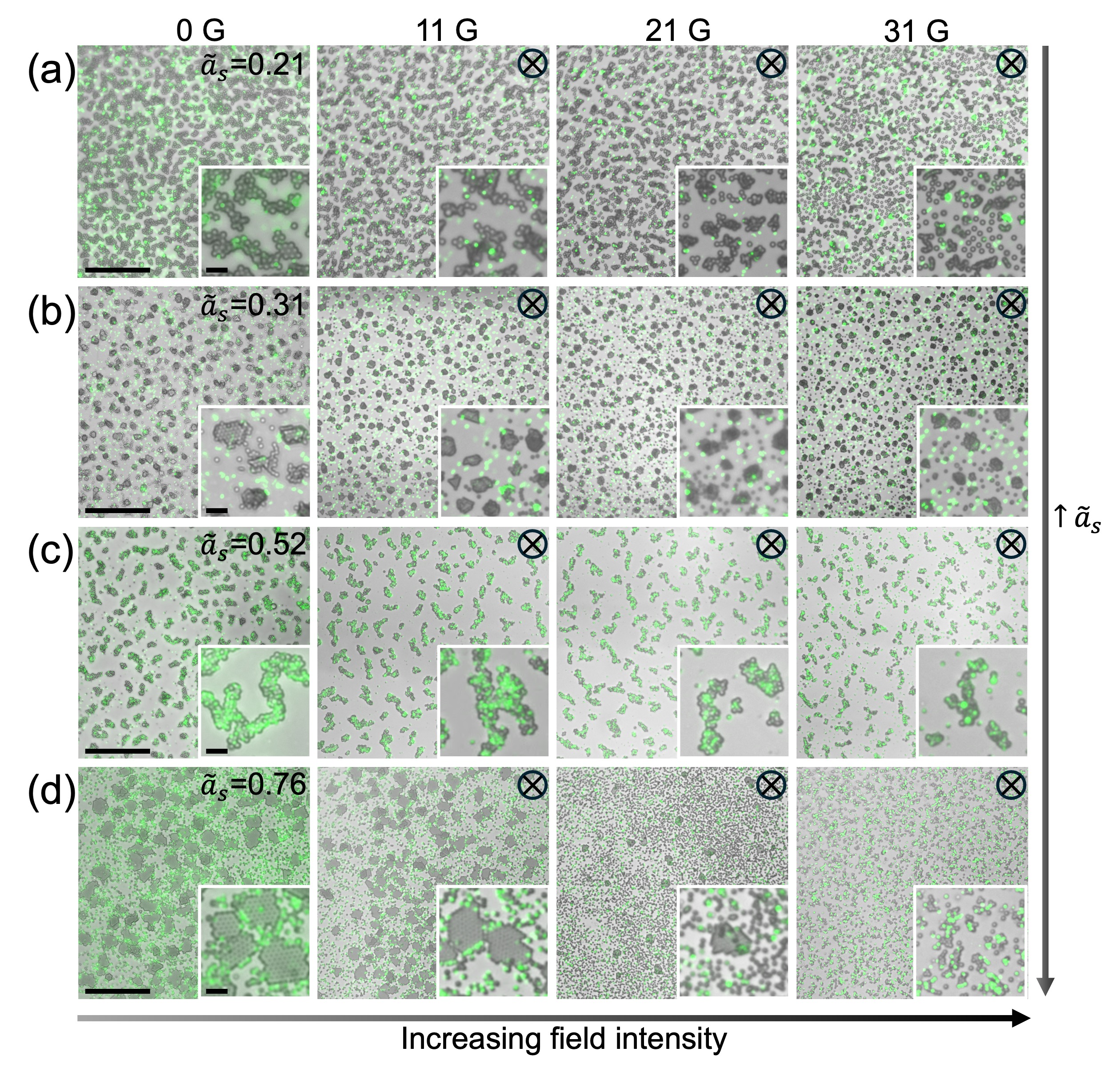}
  \caption{\textbf{Effect of particle size ratio and magnetic field strength on cluster persistence in binary colloidal monolayers.} Micrographs show $\tilde{a}_\text{S}$ of (a) 0.21, (b) 0.31, (c) 0.52, and (d) 0.76 at applied magnetic fields of 0, 11, 21, and 31 G. Among the ratios tested, clusters persist most strongly at $\tilde{a}_\text{S}=0.52$ and remain assembled under the applied field strength of 31 G. Scale bars in the main panels represent 80~\ensuremath{\mu}m and 10~\ensuremath{\mu}m in the insets.}
  \label{fig:experimental-field-response}
\end{figure}

After the initial assemblies formed, we applied a uniform magnetic field normal to the monolayer using a custom-built air-core solenoid. We adjusted the field strength by varying the current supplied to the solenoid with a DC power supply (Agilent E3611A) and measured it with a gaussmeter (AlphaLab GM2). We imaged the assemblies and their field-driven transformation in-situ using a Leica DM6 upright microscope equipped with a DFC9000 GTC camera and an EL 6000 fluorescence illuminator. We used fluorescence and bright-field microscopy to monitor the assemblies at field strengths of 0, 11, 21, and 31 G. The optical micrographs show that cluster retention as the field increased varied nonmonotonically with $\tilde{a}_\text{S}$ (Figure~\ref{fig:experimental-field-response}). At $\tilde{a}_\text{S}=0.21$ and 0.31, smaller 2D colloidal clusters became increasingly evident at higher fields. In contrast, the mixture at $\tilde{a}_\text{S}=0.52$ showed the strongest cluster resilience to magnetic field strength among the four ratios investigated, with large clusters containing both particle sizes still clearly visible at 31 G. At $\tilde{a}_\text{S}=0.76$, the extended cluster domains visible at 0 and 11 G largely broke up by 21 G, leaving dispersed particles and small crystals at 31 G. 
The persistence of binary clusters at $\tilde{a}_\text{S}=0.52$ is consistent with the simulated trends of Figure~\ref{fig:bd_snapshots_custom_grid} and the maximum in the simulated characteristic cluster persistence time at $\tilde{a}_\text{S}=0.50$ (Figure~\ref{fig:fsac}). Thus, the experiments and simulations indicate that an intermediate size ratio of $\tilde{a}_\text{S}=0.50$ favors mixed-species clusters that persist upon the application of strong magnetic fields, highlighting fundamental and nontrivial changes in the nature of interparticle interactions in the binary monolayers with microsphere size ratio.

\section{Conclusions}

This study examines quasi-two-dimensional, binary monolayers of nonmagnetic microspheres immersed in a ferrofluid as a platform for in situ tuning of nonequilibrium colloidal organization and dynamics. Specifically, the focus is on understanding how particle-size asymmetry, mixing ratio, and strength of a normally applied magnetic field modify the assembly and dynamic persistence of microsphere clusters. The experiments and the model investigated via BD simulations establish that the magnetic field's effects on cluster formation vary nonmonotonically with microsphere radius ratio, with a strong preference for cluster formation evident at a 2:1 microsphere radius ratio. Due to gravity, the large and small microsphere centers do not reside in the same plane, giving rise to non-additive hard-sphere interactions. As a result, a 2:1 radius ratio maps to an effective ratio of 2.414:1, the magic ratio for the S1 square binary alloy where contact attractions are maximized.
Dynamic cluster persistence of the simulated model monolayers were also found to vary nonmonotonically with microsphere size ratio, with the slowest colloid exchange rates occurring at the same magic ratio. Structural analysis across all of the monolayers investigated showed that, when specific crystalline motifs that maximize interparticle contact attractions emerged ($\mathrm{HEX_{\text{L}}}$, S1, and T1), the cluster persistence time correlated with the fraction of microspheres contributing to those motifs. Interestingly, cluster exchange dynamics were also shown to strongly correlate to the Ollivier-Ricci curvature of force-magnitude weighted graphs describing the microsphere clusters, irrespective of the specific crystalline motifs present.   

In future studies, it will be interesting to learn whether Ollivier-Ricci curvature, which has been recently advanced as a measure of complexity in colloidal assemblies that combine order and disorder,\cite{hallstrom2026decoding} more broadly correlates with rates of dynamic reorganization in dense, many-body interaction networks.\cite{yeh2026dynamic} 
Because the underlying graph construction requires only knowledge of pairwise interaction strengths, the same framework could in principle be extended to understand dynamic relaxation of other size- or shape-disperse colloidal mixtures, systems with additional interaction types (e.g., combined paramagnetic/diamagnetic mixtures), or more complex materials such as biological aggregates such as cell clusters. Coupling such curvature-based descriptors with data-driven feature-importance analysis, as is done here, is a step toward inverse design.\cite{sherman2020inverse,dijkstra2021predictive} Identifying which structural or interaction parameters most strongly control aggregate dynamics could guide the choice of size ratio, composition, or time-varying field protocols\cite{swan2014directed,sherman2016dynamic,spatafora2021hierarchical,conradt2026dissipative} needed to target a desired persistence or relaxation behavior in out-of-equilibrium colloidal assemblies.

\section{Supplemental Material}
See the supporting information for additional simulation
details, including additional results for ORC of force-magnitude weighted graphs, FSAC, binary alloy structural motifs, and particle size distributions.

\section{Acknowledgments}
We acknowledge the Texas Advanced Computing Center (TACC) at The University of Texas at Austin for providing HPC resources. This work was supported in part by NSF grant 2323482 and by the “Center of Complex Particle Systems (COMPASS)” (grant NSF 2243104). All experimental work for observations of local packing and cluster persistence were supported by the U.S. Department of Energy (DOE), Office of Science, Basic Energy Sciences (BES), Materials Sciences and Engineering, Synthesis and Processing Science (S\&PS) program under Grant No. DE-SC0026094.

\section{Author Declarations}
\subsection{Conflict of Interest}

The authors have no conflicts to disclose.

\subsection{Author Contributions}
Conceptualization: C.C.G.Y., J.G.L., P.B., B.B., T.M.T. Investigation: C.C.G.Y., J.G.L.. Writing -- review \& editing: C.C.G.Y., T.M.T. Methodology: C.C.G.Y., J.G.L., P.B., B.B., T.M.T. Resources: B.B., T.M.T. Funding acquisition:  P.B., B.B., T.M.T. Supervision:  B.B., T.M.T. Data curation:  C.C.G.Y.,J.G.L. Validation:  C.C.G.Y., J.G.L., P.B., B.B., T.M.T. Formal analysis:  C.C.G.Y., J.G.L. Project administration:  B.B., T.M.T. Software:  C.C.G.Y.. Visualization:  C.C.G.Y., J.G.L., B.B., T.M.T.

\section{Data Availability}
All data and code needed to evaluate and reproduce the results in the paper are present in the paper and/or the Supplementary Materials.

\renewcommand\refname{}\vspace{-10mm}
\bibliography{references}
\newpage

\noindent \center \title{\LARGE{Supporting Information for \\  
Size asymmetry and dynamic cluster persistence in quasi-two-dimensional colloids with competitive interactions
}}


\maketitle

\begin{center}
\noindent
\small
\textbf{*Corresponding authors:} bbharti@lsu.edu (B. B.); truskett@umich.edu (T.M.T.)
\end{center}


\renewcommand{\thefigure}{S\arabic{figure}}
\setcounter{figure}{0}

\begin{figure}[!htbp]
    \centering
    \includegraphics[width=1\linewidth]{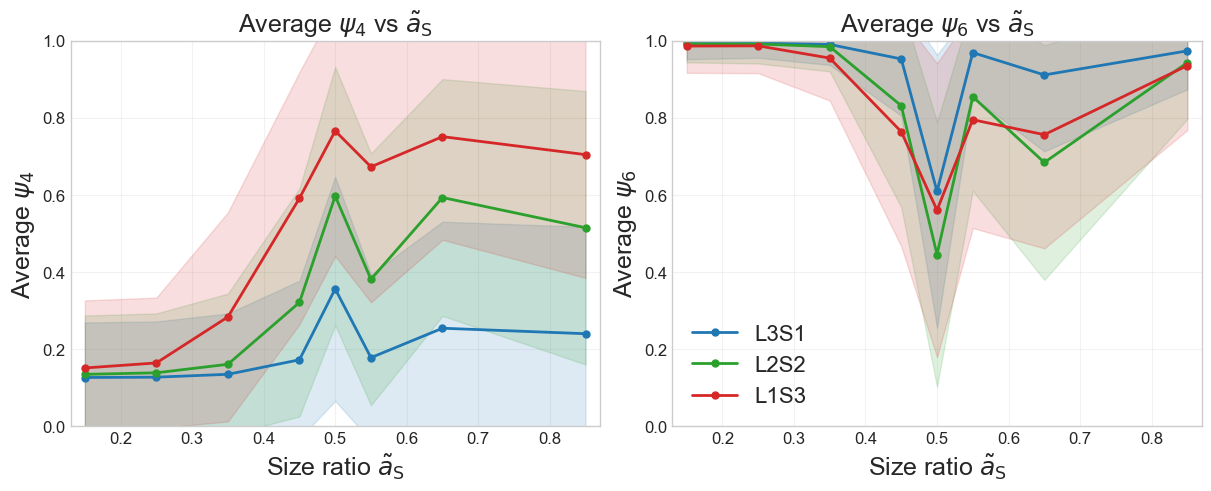}
    \caption{{\bf{Mean local bond-orientational order parameters as a function of size ratio in the binary mixture.}} Left panel: $\langle \psi_4 \rangle$ vs.\ $\tilde{a}_\text{S}$; right panel: $\langle \psi_6 \rangle$ vs.\ $\tilde{a}_\text{S}$. Curves correspond to composition groups $x_\text{S}=0.25$ (L3S1, blue), $x_\text{S}=0.50$ (L2S2, green), and $x_\text{S}=0.75$ (L1S3, red). Symbols show sample means computed from node-level (i.e., particle-level) values extracted from the dataset, and shaded bands indicate one standard deviation across sampled nodes within each $(\tilde a_\text{S}, x_\text{S})$ state. 
    The trends quantify how fourfold and sixfold local order evolve with particle size asymmetry and mixing ratio.}
    \label{fig:psisizeratio}
\end{figure}

\begin{figure}
    \centering
    \includegraphics[width=1\linewidth]{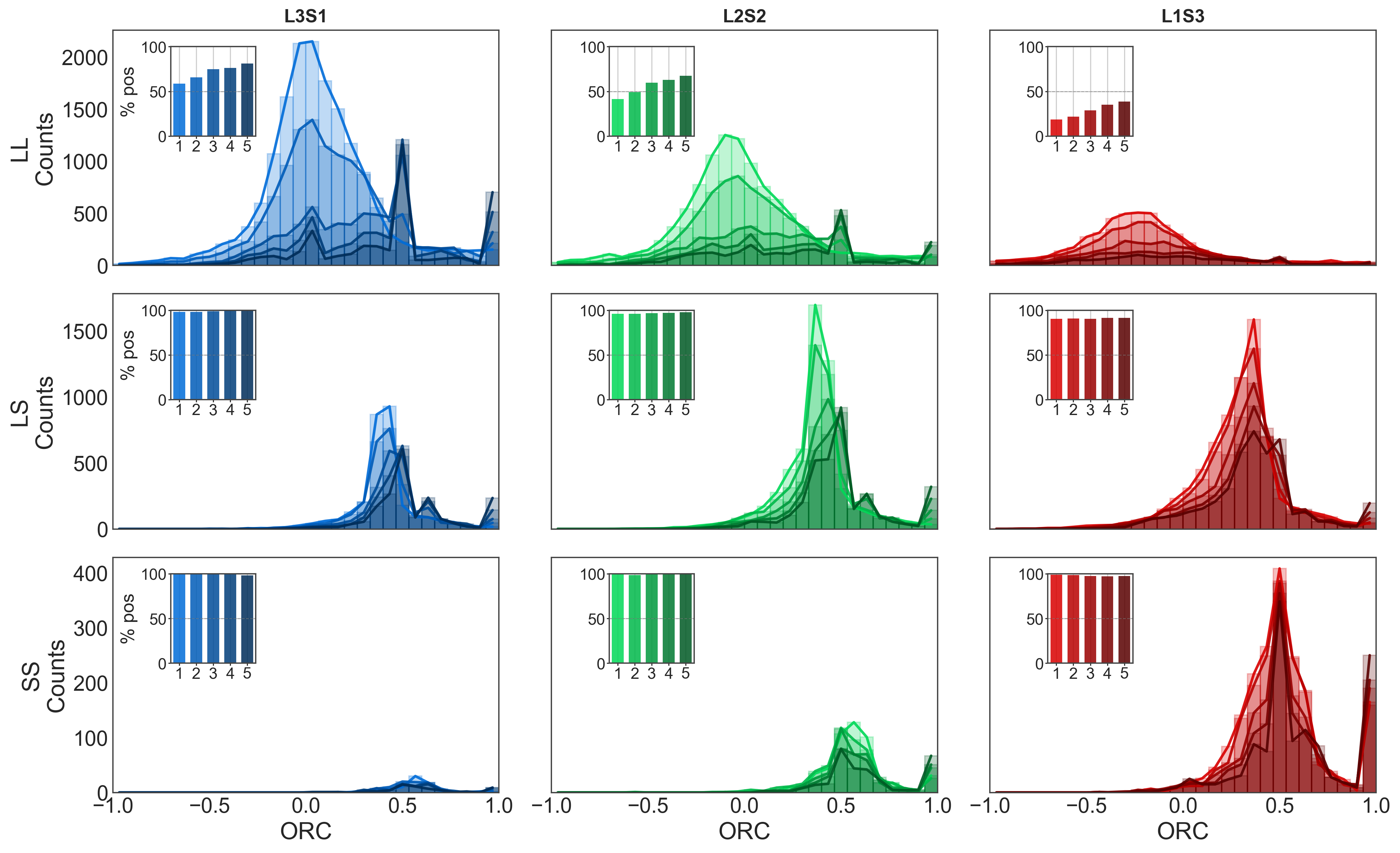}
    \caption{Ollivier-Ricci curvature distribution function for edges by absolute count of BD simulated configurations at \(\tilde{\varepsilon}_\text{d}=7.5\) and $\tilde a_\text{S}=0.25$. From top to bottom, rows present data for LL, LS, SS edges.  From left to right, results are for monolayers with different L:S microsphere mixing ratios:  L3S1 (blue), L2S2 (green), and L1S3 (red). From bright to dark, the colored curves represent \(\tilde{\varepsilon}_\text{m}=1,2,3,4,5\).}
    \label{fig:placeholder}
\end{figure}

\begin{figure}
    \centering
    \includegraphics[width=1\linewidth]{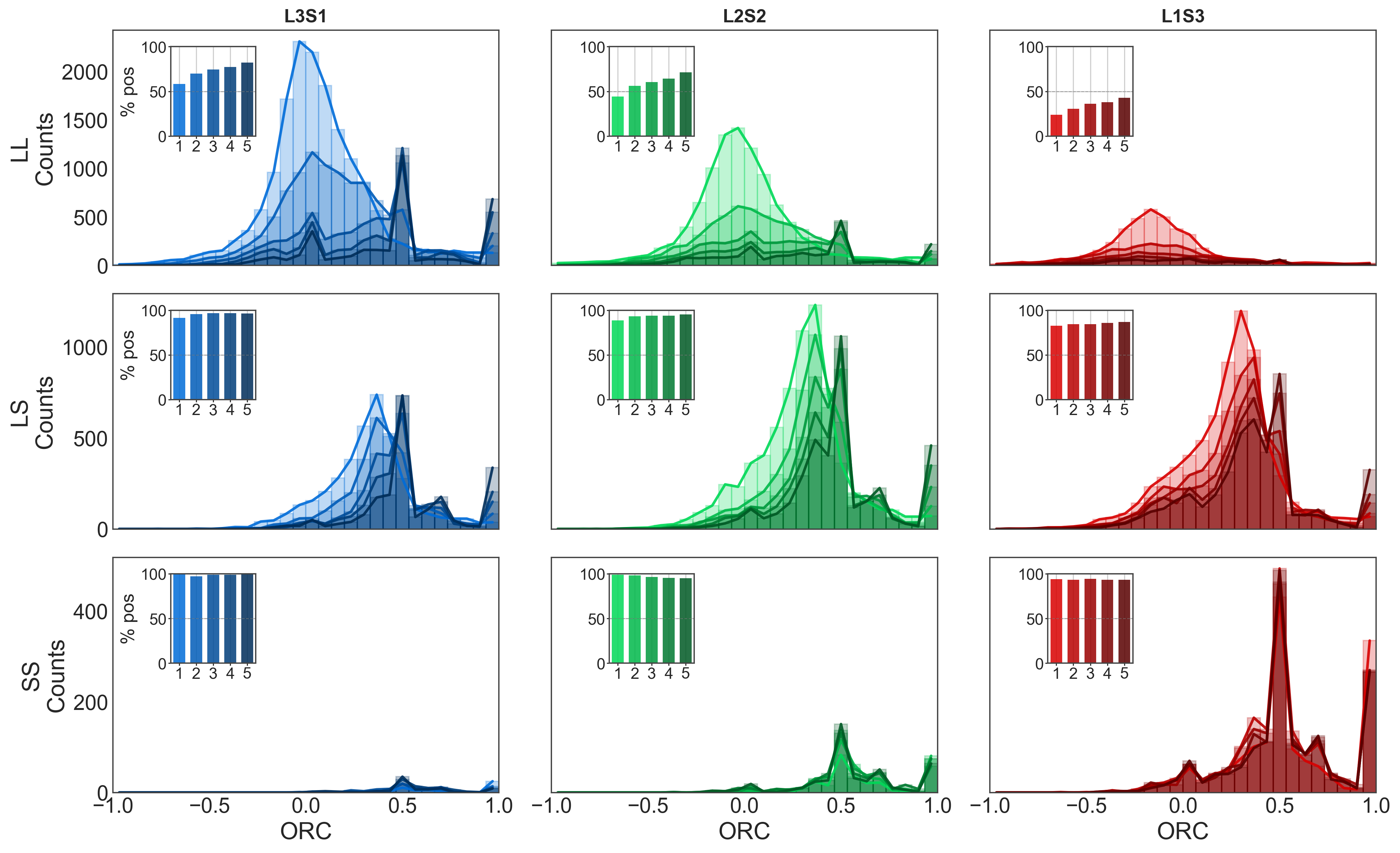}
    \caption{Ollivier-Ricci curvature distribution function for edges by absolute count of BD simulated configurations at \(\tilde{\varepsilon}_\text{d}=7.5\) and $\tilde a_\text{S}=0.35$. From top to bottom, rows present data for LL, LS, SS edges.  From left to right, results are for monolayers with different L:S microsphere mixing ratios:  L3S1 (blue), L2S2 (green), and L1S3 (red). From bright to dark, the colored curves represent \(\tilde{\varepsilon}_\text{m}=1,2,3,4,5\).}
    \label{fig:placeholder}
\end{figure}

\begin{figure}
    \centering
    \includegraphics[width=1\linewidth]{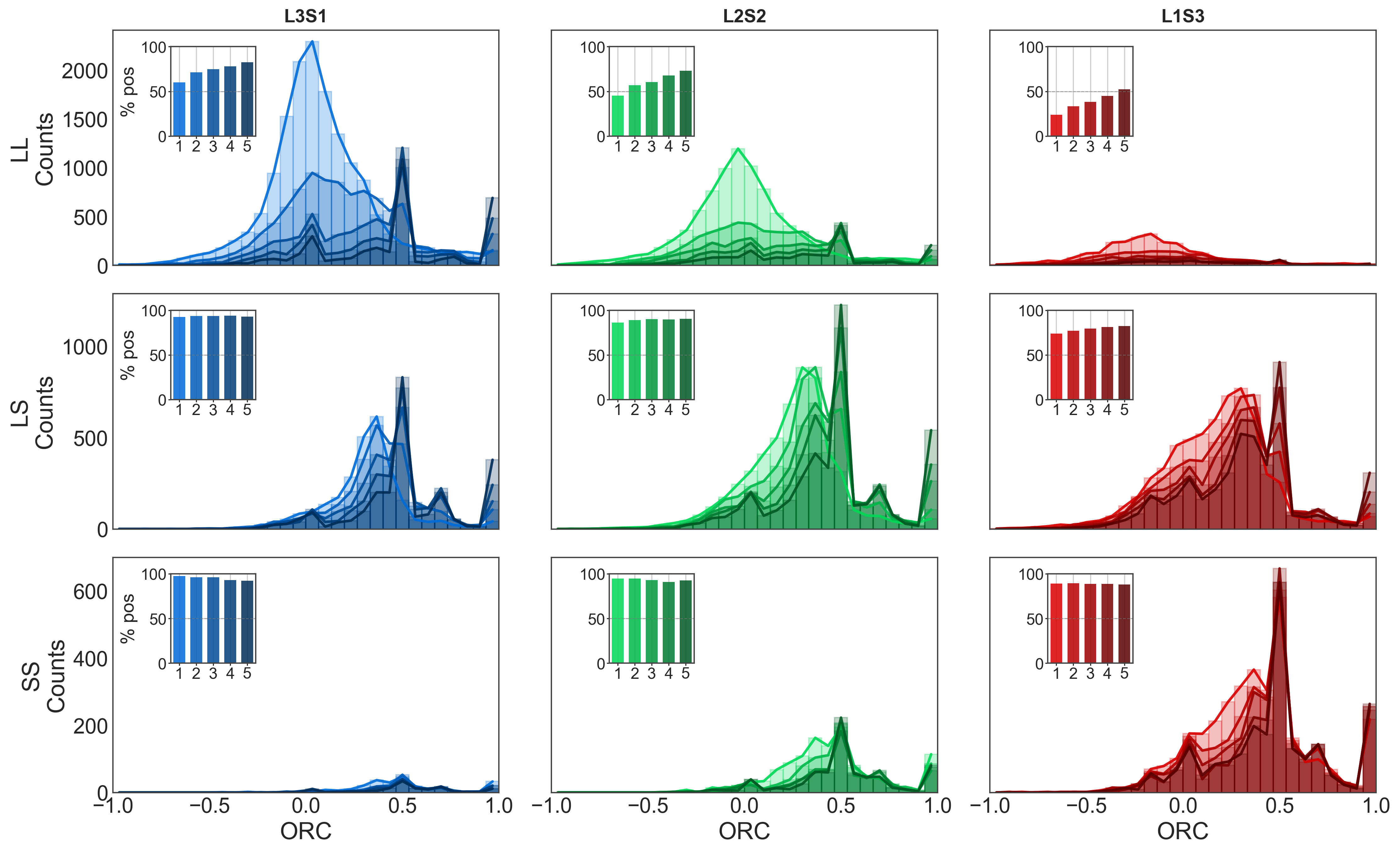}
    \caption{Ollivier-Ricci curvature distribution function for edges by absolute count of BD simulated configurations at \(\tilde{\varepsilon}_\text{d}=7.5\) and $\tilde a_\text{S}=0.45$. From top to bottom, rows present data for LL, LS, SS edges.  From left to right, results are for monolayers with different L:S microsphere mixing ratios:  L3S1 (blue), L2S2 (green), and L1S3 (red). From bright to dark, the colored curves represent \(\tilde{\varepsilon}_\text{m}=1,2,3,4,5\).}
    \label{fig:placeholder}
\end{figure}

\begin{figure}
    \centering
    \includegraphics[width=1\linewidth]{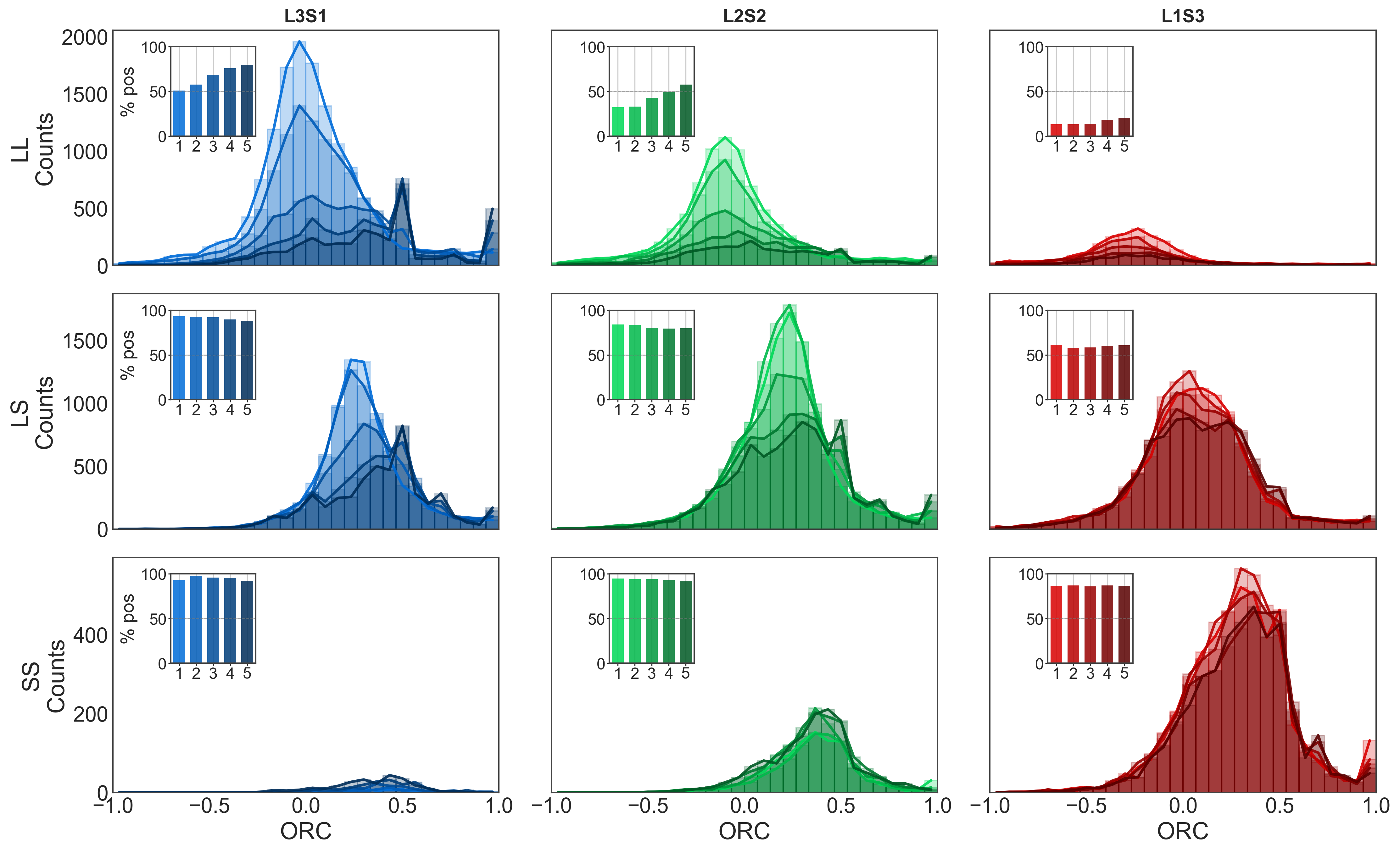}
    \caption{Ollivier-Ricci curvature distribution function for edges by absolute count of BD simulated configurations at \(\tilde{\varepsilon}_\text{d}=7.5\) and $\tilde a_\text{S}=0.5$. From top to bottom, rows present data for LL, LS, SS edges.  From left to right, results are for monolayers with different L:S microsphere mixing ratios:  L3S1 (blue), L2S2 (green), and L1S3 (red). From bright to dark, the colored curves represent \(\tilde{\varepsilon}_\text{m}=1,2,3,4,5\).}
    \label{fig:placeholder}
\end{figure}

\begin{figure}
    \centering
    \includegraphics[width=1\linewidth]{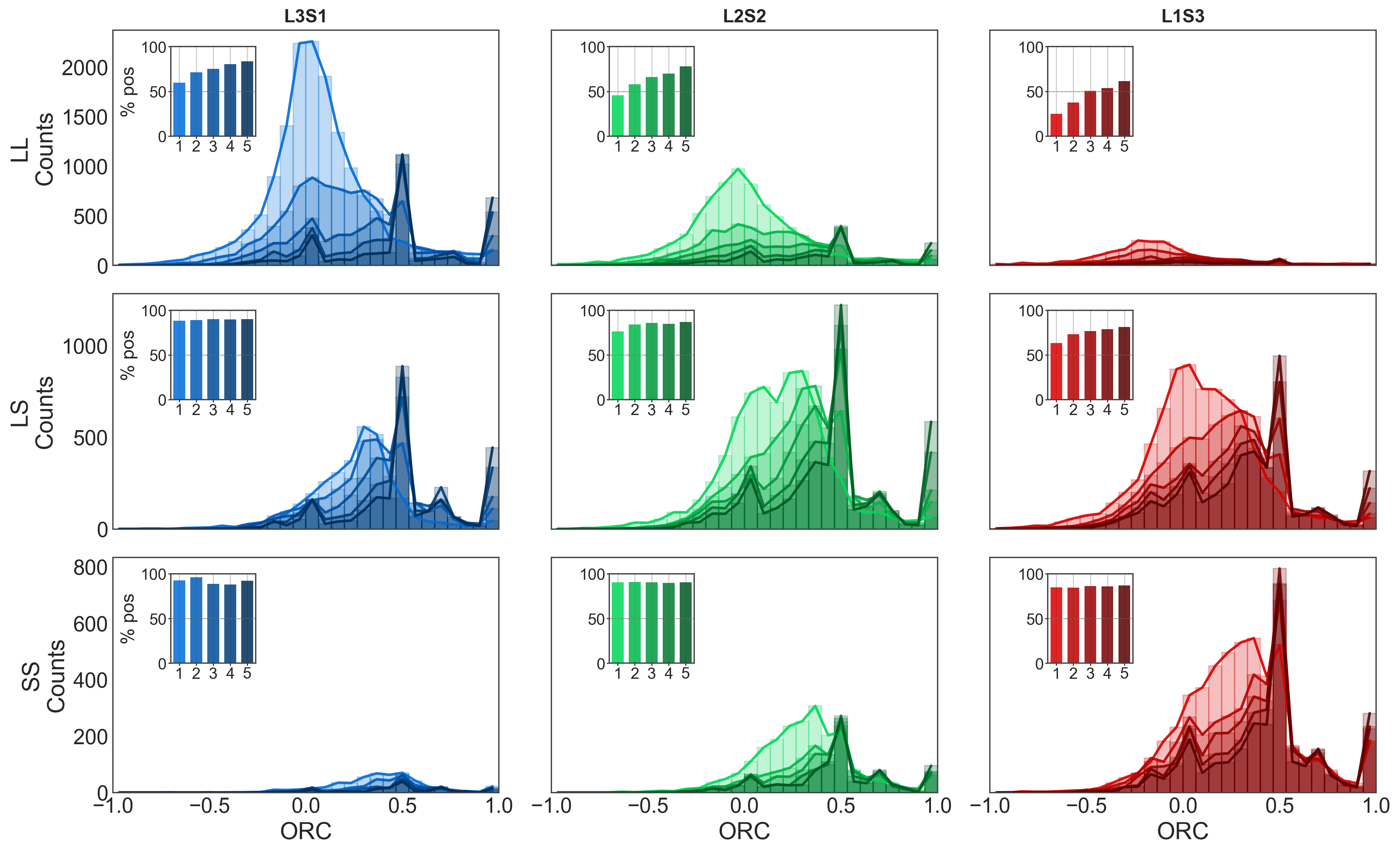}
    \caption{Ollivier-Ricci curvature distribution function for edges by absolute count of BD simulated configurations at \(\tilde{\varepsilon}_\text{d}=7.5\) and $\tilde a_\text{S}=0.55$. From top to bottom, rows present data for LL, LS, SS edges.  From left to right, results are for monolayers with different L:S microsphere mixing ratios:  L3S1 (blue), L2S2 (green), and L1S3 (red). From bright to dark, the colored curves represent \(\tilde{\varepsilon}_\text{m}=1,2,3,4,5\).}
    \label{fig:placeholder}
\end{figure}

\begin{figure}
    \centering
    \includegraphics[width=1\linewidth]{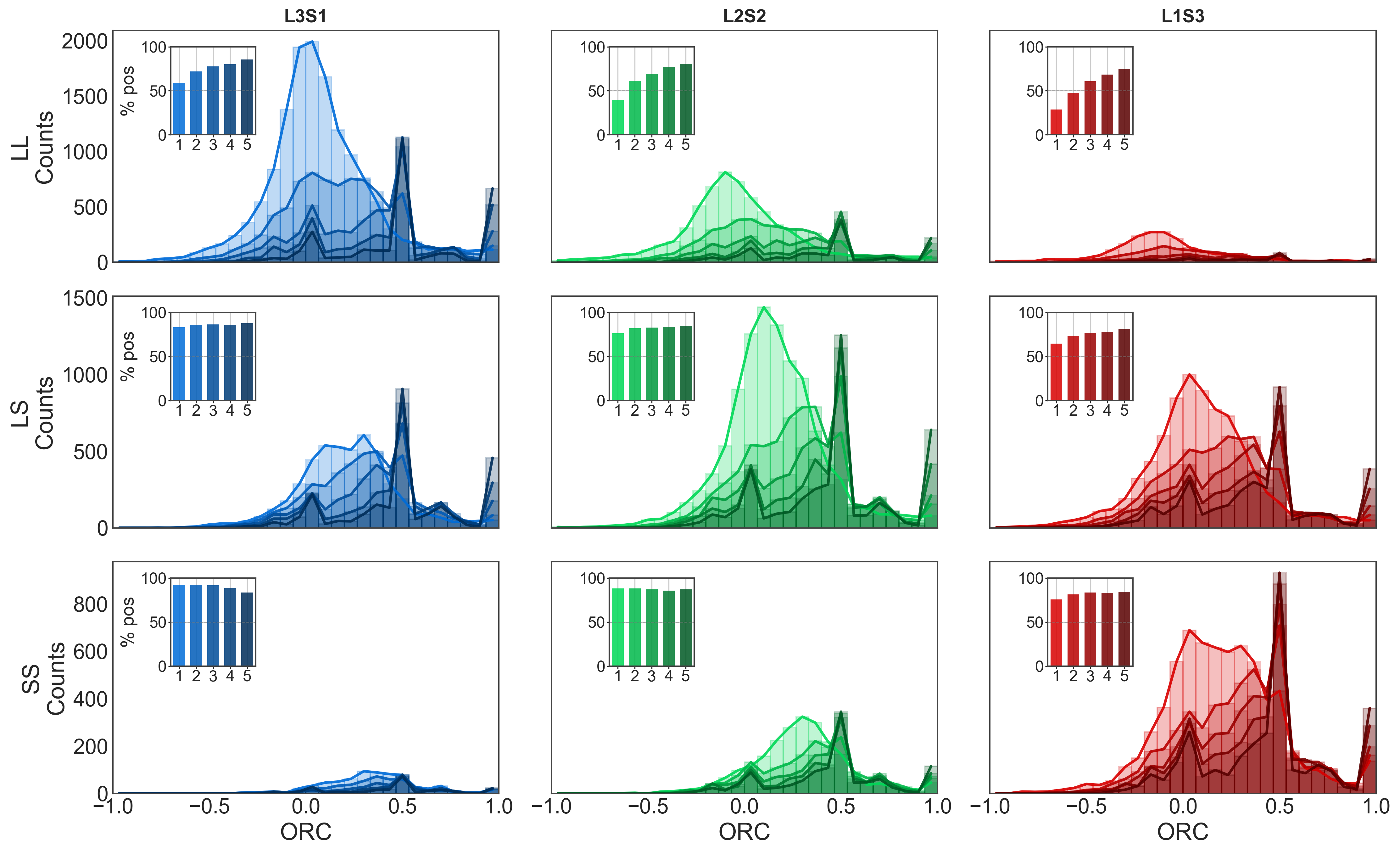}
    \caption{Ollivier-Ricci curvature distribution function for edges by absolute count of BD simulated configurations at \(\tilde{\varepsilon}_\text{d}=7.5\) and $\tilde a_\text{S}=0.65$. From top to bottom, rows present data for LL, LS, SS edges.  From left to right, results are for monolayers with different L:S microsphere mixing ratios:  L3S1 (blue), L2S2 (green), and L1S3 (red). From bright to dark, the colored curves represent \(\tilde{\varepsilon}_\text{m}=1,2,3,4,5\).}
    \label{fig:placeholder}
\end{figure}

\begin{figure}
    \centering
    \includegraphics[width=1\linewidth]{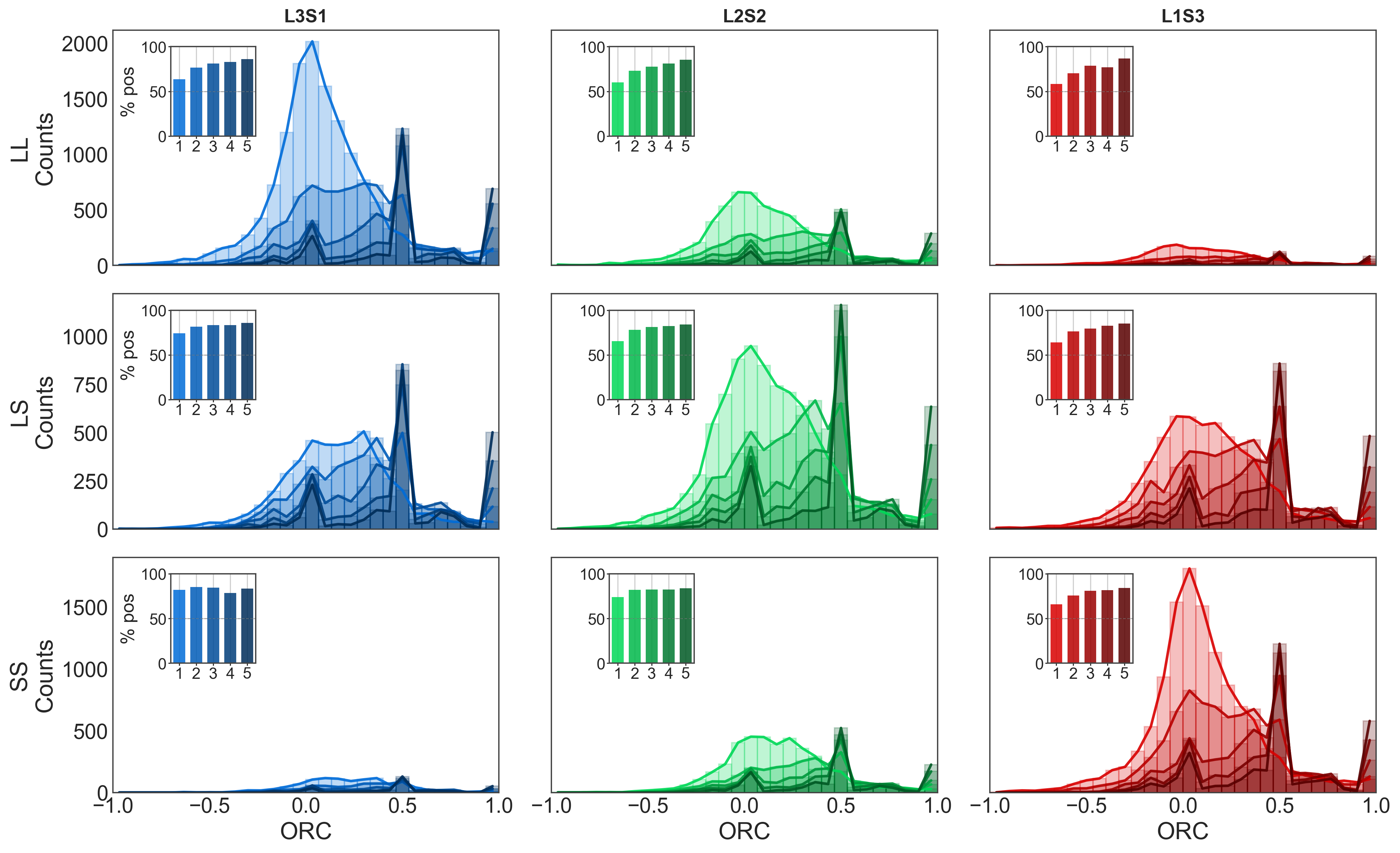}
    \caption{Ollivier-Ricci curvature distribution function for edges by absolute count of BD simulated configurations at \(\tilde{\varepsilon}_\text{d}=7.5\) and $\tilde a_\text{S}=0.85$. From top to bottom, rows present data for LL, LS, SS edges.  From left to right, results are for monolayers with different L:S microsphere mixing ratios:  L3S1 (blue), L2S2 (green), and L1S3 (red). From bright to dark, the colored curves represent \(\tilde{\varepsilon}_\text{m}=1,2,3,4,5\).}
    \label{fig:placeholder}
\end{figure}



\begin{figure}
    \centering
    \includegraphics[width=0.8\linewidth]{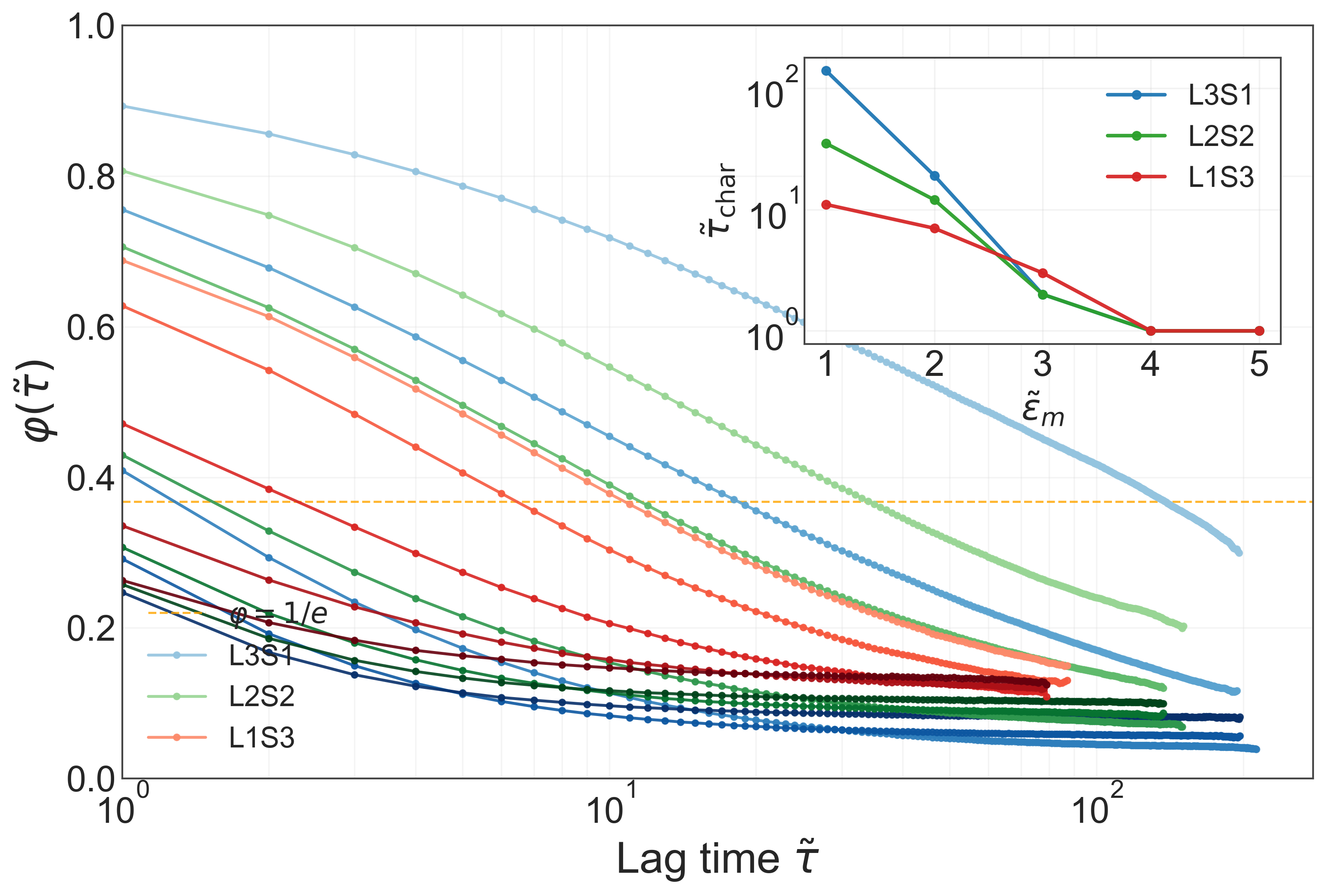}
    \caption{FSAC ($\varphi(\tilde \tau)$) quantifying cluster exchange dynamics. Results are shown for L3S1 (blue), L2S2 (green) and L1S3 (red) monolayers at ($\tilde{\varepsilon}_\text{d}=7.5$) at $\tilde a_\text{S}=0.25$. Curves from light to dark correspond to increasing dipolar interaction strength, ($\tilde{\varepsilon}_\text{m}=1–5$). Inset: characteristic lag time ($\tilde \tau_\text{char}$) defined by intersection where ($\varphi(\tilde \tau_\text{char})=e^{-1}$).}
    \label{fig:placeholder}
\end{figure}

\begin{figure}
    \centering
    \includegraphics[width=0.8\linewidth]{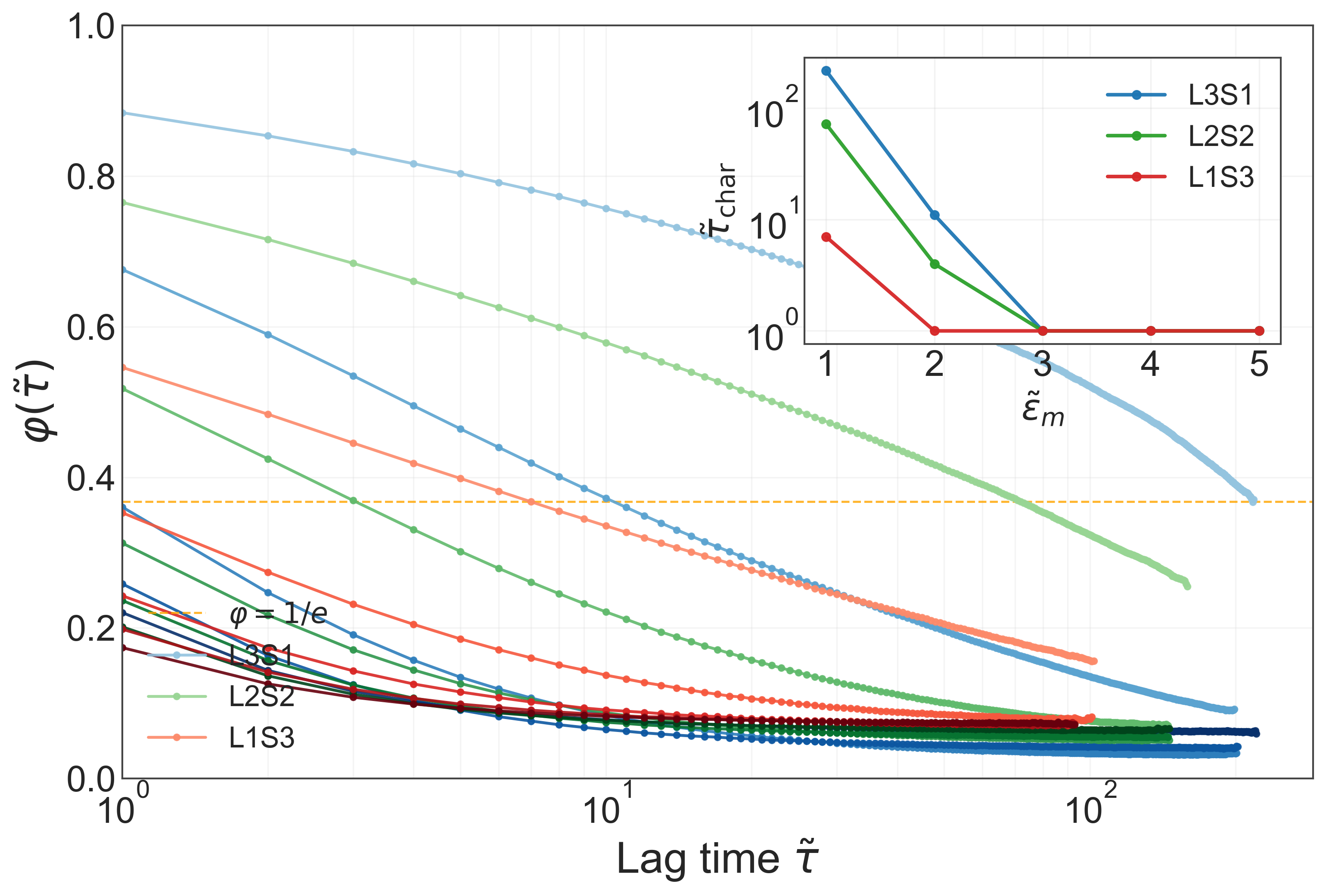}
    \caption{FSAC ($\varphi(\tilde \tau)$) quantifying cluster exchange dynamics. Results are shown for L3S1 (blue), L2S2 (green) and L1S3 (red) monolayers at ($\tilde{\varepsilon}_\text{d}=7.5$) at $\tilde a_\text{S}=0.35$. Curves from light to dark correspond to increasing dipolar interaction strength, ($\tilde{\varepsilon}_\text{m}=1–5$). Inset: characteristic lag time ($\tilde \tau_\text{char}$) defined by intersection where ($\varphi(\tilde \tau_\text{char})=e^{-1}$).}
    \label{fig:placeholder}
\end{figure}

\begin{figure}
    \centering
    \includegraphics[width=0.8\linewidth]{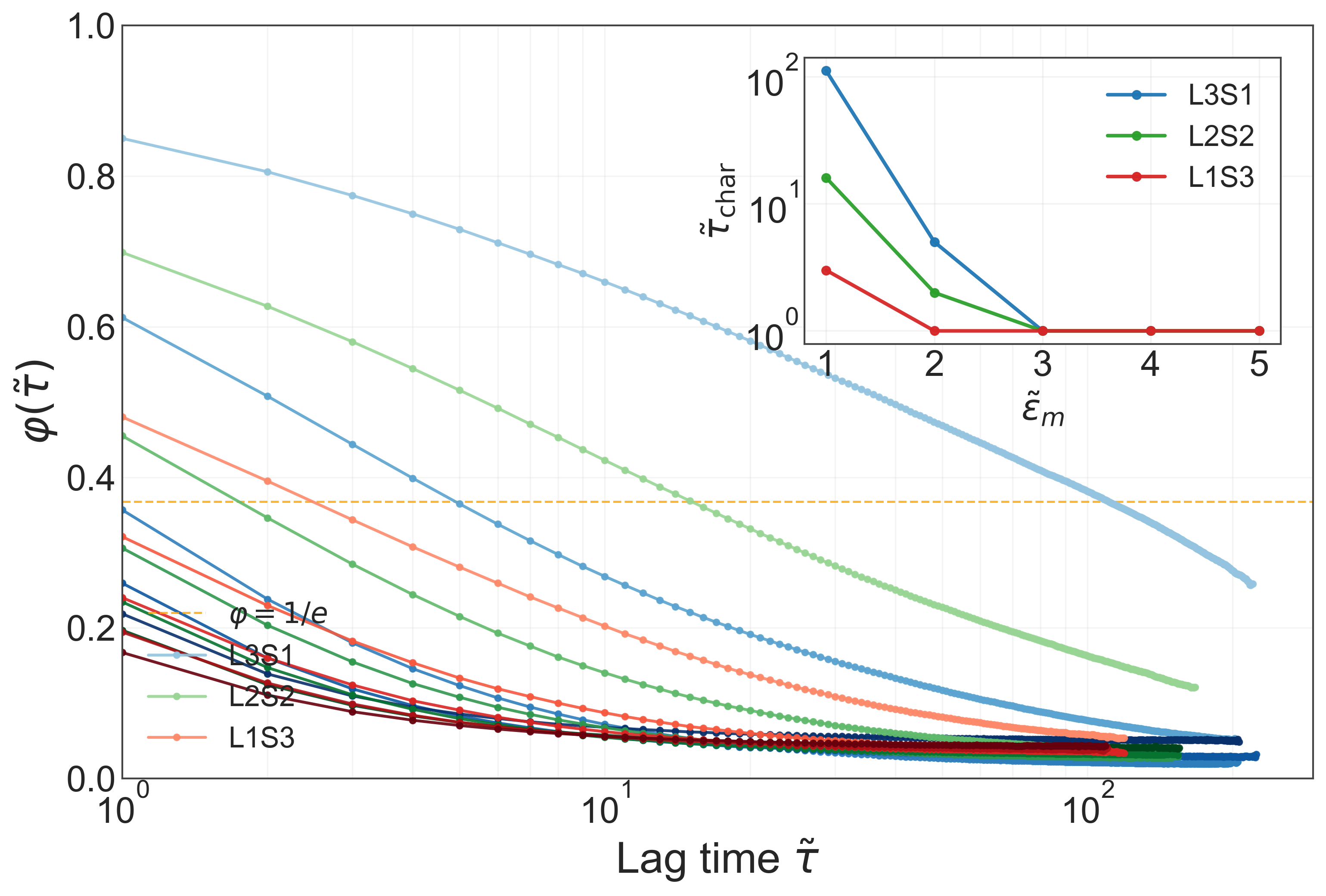}
    \caption{FSAC ($\tilde \varphi(\tau)$) quantifying cluster exchange dynamics. Results are shown for L3S1 (blue), L2S2 (green) and L1S3 (red) monolayers at ($\tilde{\varepsilon}_\text{d}=7.5$) at $\tilde a_\text{S}=0.45$. Curves from light to dark correspond to increasing dipolar interaction strength, ($\tilde{\varepsilon}_\text{m}=1–5$). Inset: characteristic lag time ($\tilde \tau_\text{char}$) defined by intersection where ($\varphi(\tilde \tau_\text{char})=e^{-1}$).}
    \label{fig:placeholder}
\end{figure}

\begin{figure}
    \centering
    \includegraphics[width=0.8\linewidth]{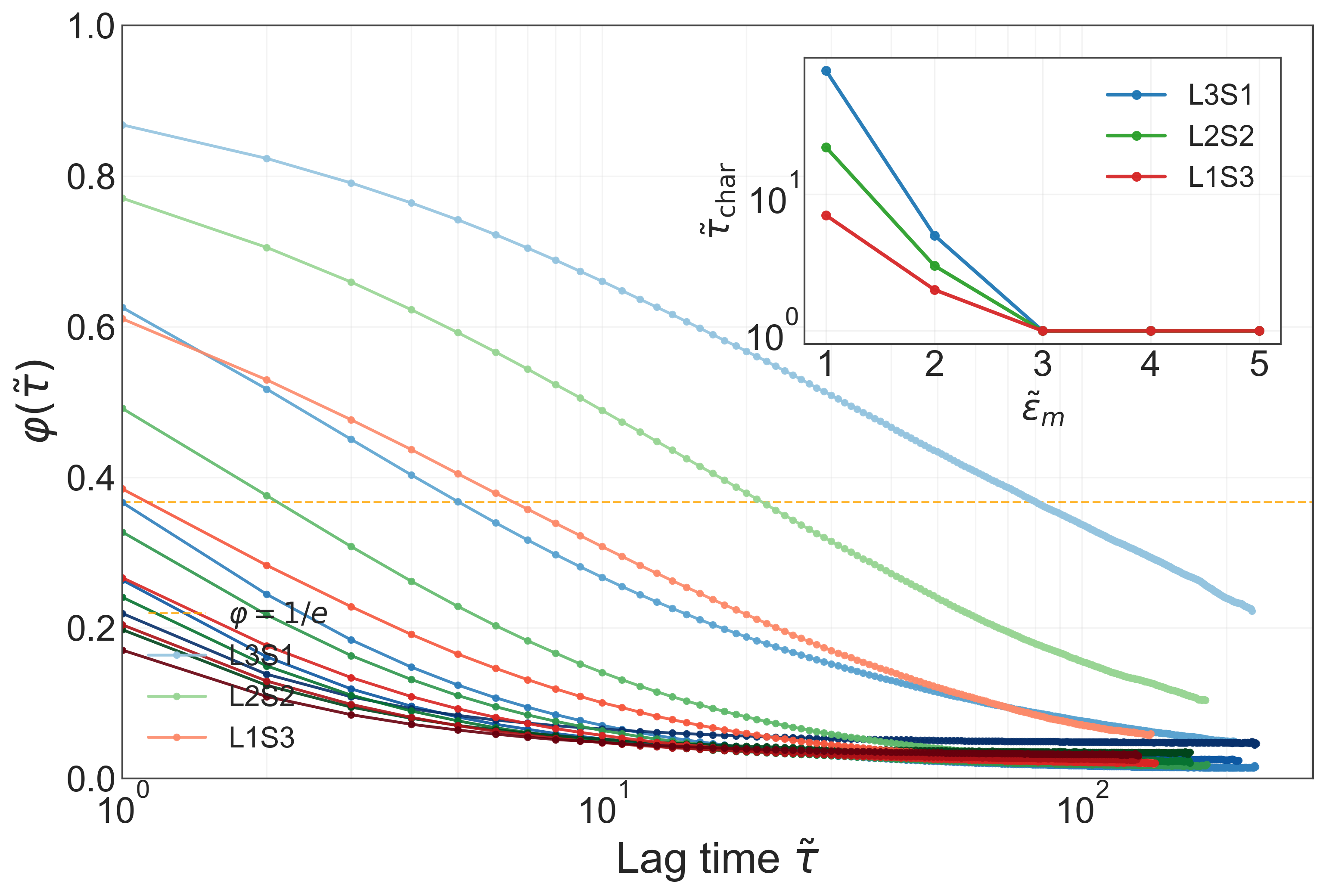}
    \caption{FSAC ($\varphi(\tilde \tau)$) quantifying cluster exchange dynamics. Results are shown for L3S1 (blue), L2S2 (green) and L1S3 (red) monolayers at ($\tilde{\varepsilon}_\text{d}=7.5$) at $\tilde a_\text{S}=0.55$. Curves from light to dark correspond to increasing dipolar interaction strength, ($\tilde{\varepsilon}_\text{m}=1–5$). Inset: characteristic lag time ($\tilde \tau_\text{char}$) defined by intersection where ($\varphi(\tilde \tau_\text{char})=e^{-1}$).}
    \label{fig:placeholder}
\end{figure}

\begin{figure}
    \centering
    \includegraphics[width=0.8\linewidth]{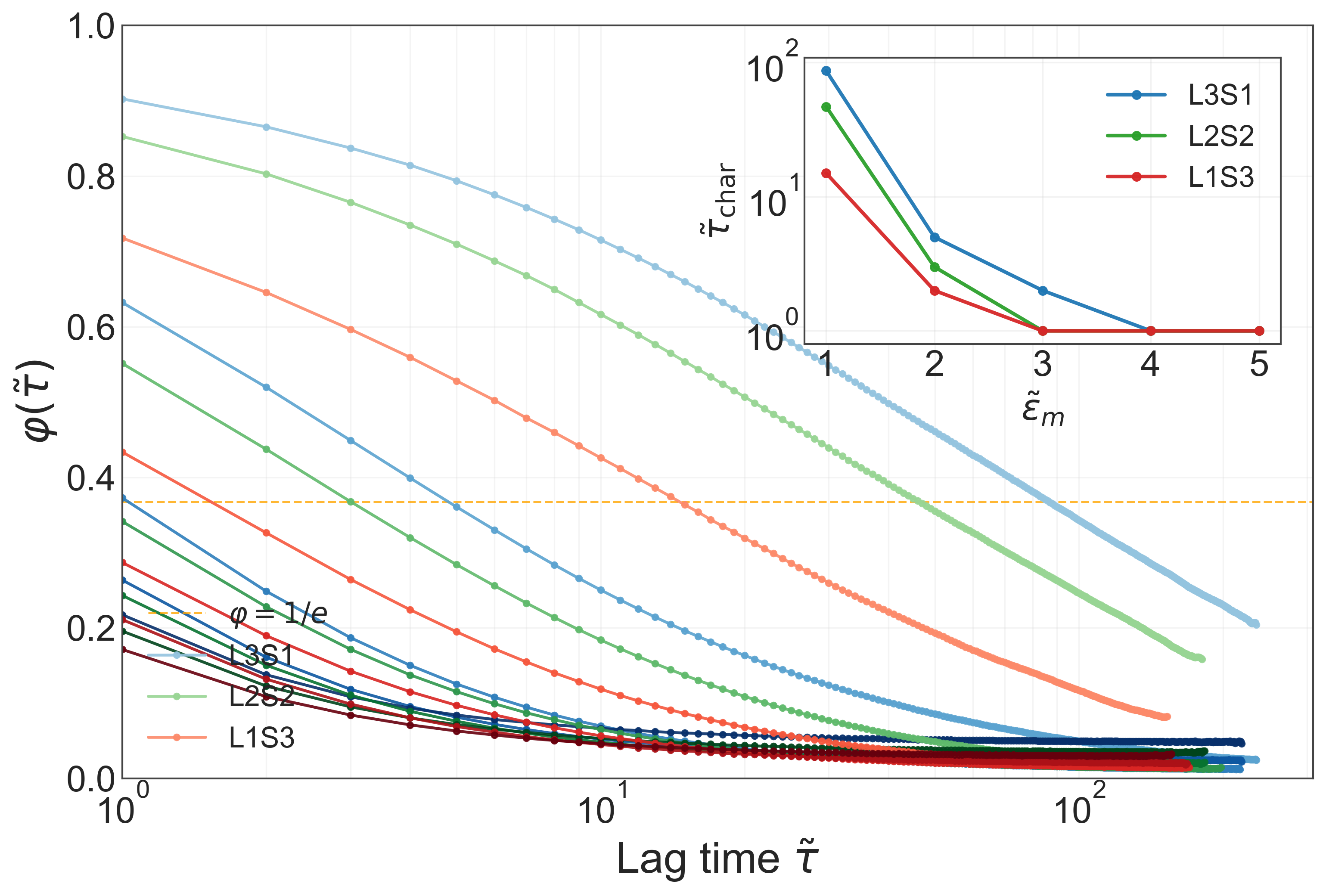}
    \caption{FSAC ($\varphi(\tilde \tau)$) quantifying cluster exchange dynamics. Results are shown for L3S1 (blue), L2S2 (green) and L1S3 (red) monolayers at ($\tilde{\varepsilon}_\text{d}=7.5$) at $\tilde a_\text{S}=0.65$. Curves from light to dark correspond to increasing dipolar interaction strength, ($\tilde{\varepsilon}_\text{m}=1–5$). Inset: characteristic lag time ($\tilde \tau_\text{char}$) defined by intersection where ($\varphi(\tilde \tau_\text{char})=e^{-1}$).}
    \label{fig:placeholder}
\end{figure}

\begin{figure}
    \centering
    \includegraphics[width=0.8\linewidth]{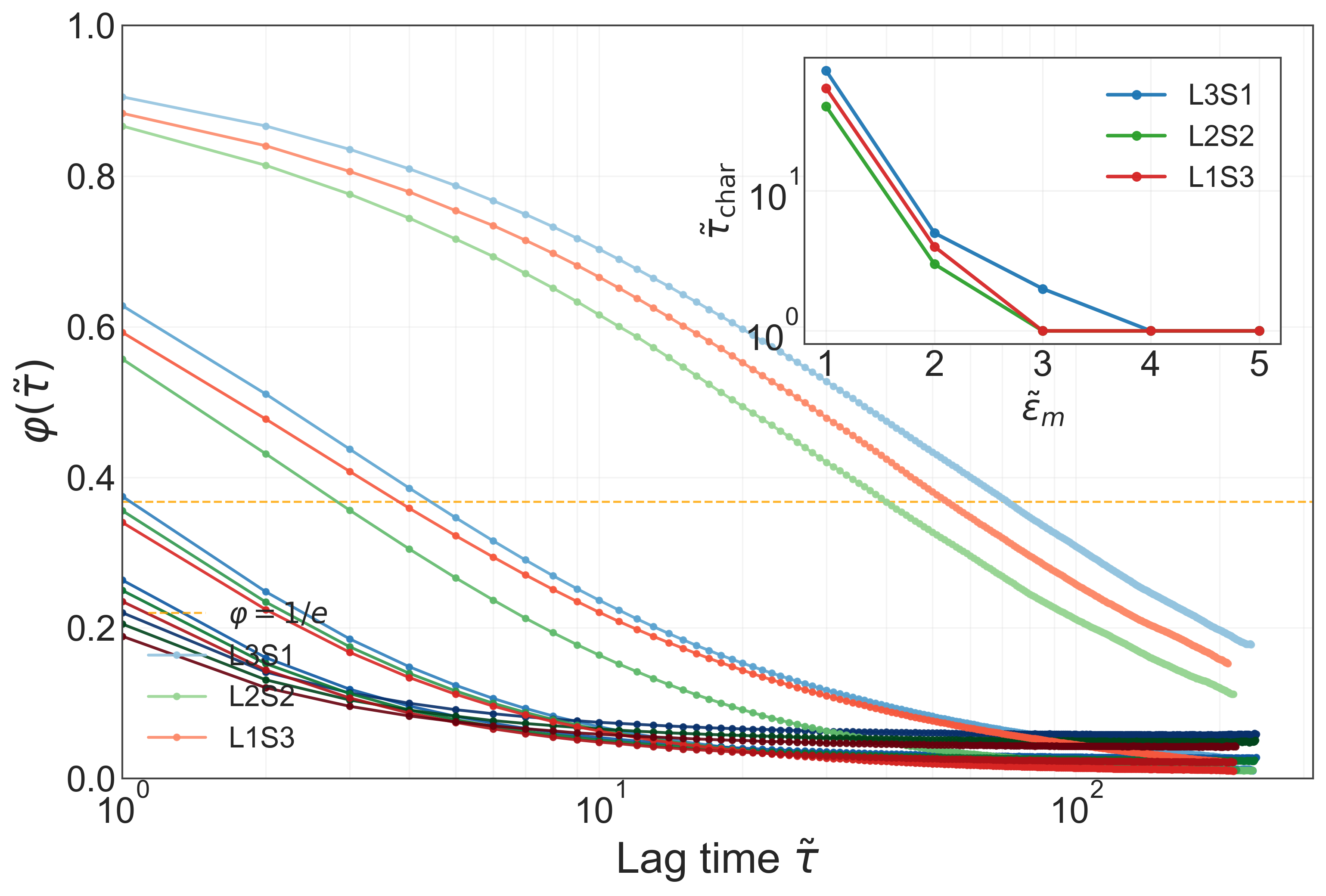}
    \caption{FSAC ($\varphi(\tilde \tau)$) quantifying cluster exchange dynamics. Results are shown for L3S1 (blue), L2S2 (green) and L1S3 (red) monolayers at ($\tilde{\varepsilon}_\text{d}=7.5$) at $\tilde a_\text{S}=0.85$. Curves from light to dark correspond to increasing dipolar interaction strength, ($\tilde{\varepsilon}_\text{m}=1–5$). Inset: characteristic lag time ($\tilde \tau_\text{char}$) defined by intersection where ($\varphi(\tilde \tau_\text{char})=e^{-1}$).}
    \label{fig:placeholder}
\end{figure}


\begin{figure}[tbp]
  \centering
  \includegraphics[width=\linewidth]{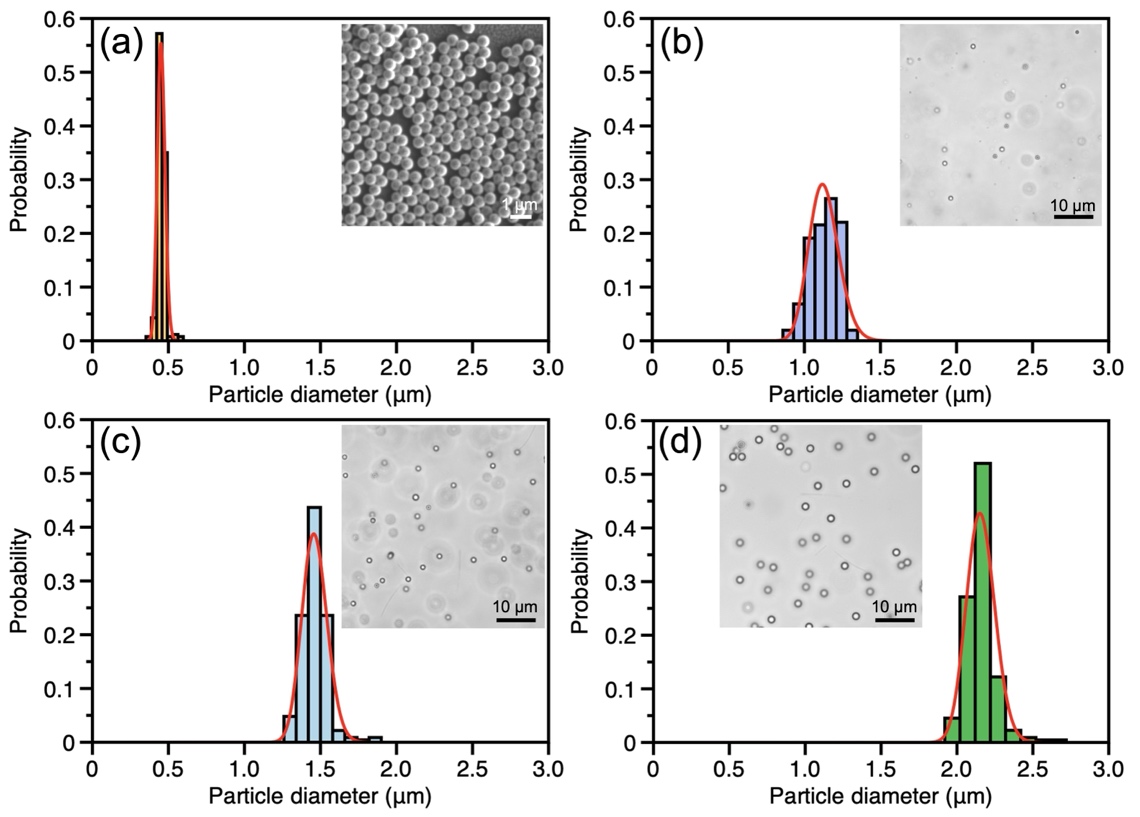}
  \caption{\textbf{Microscopy images and particle diameter distributions for the model particles.} The mean particle diameters are (a) 0.47 \ensuremath{\pm} 0.02~\ensuremath{\mu}m, (b) 1.15 \ensuremath{\pm} 0.09~\ensuremath{\mu}m, (c) 1.50 \ensuremath{\pm} 0.08~\ensuremath{\mu}m, and (d) 2.21 \ensuremath{\pm} 0.15~\ensuremath{\mu}m. Diameters were measured by SEM for the submicrometer particles in (a) and by optical microscopy for the microparticles in (b--d). Histograms show the measured diameter distributions, and solid curves show the best-fit lognormal distributions. Insets show representative SEM (a) and optical microscopy (b--d) images used for the diameter measurements.}
  \label{fig:experimental-particle-sizes}
\end{figure}

\begin{figure}
    \centering
    \includegraphics[width=1\linewidth]{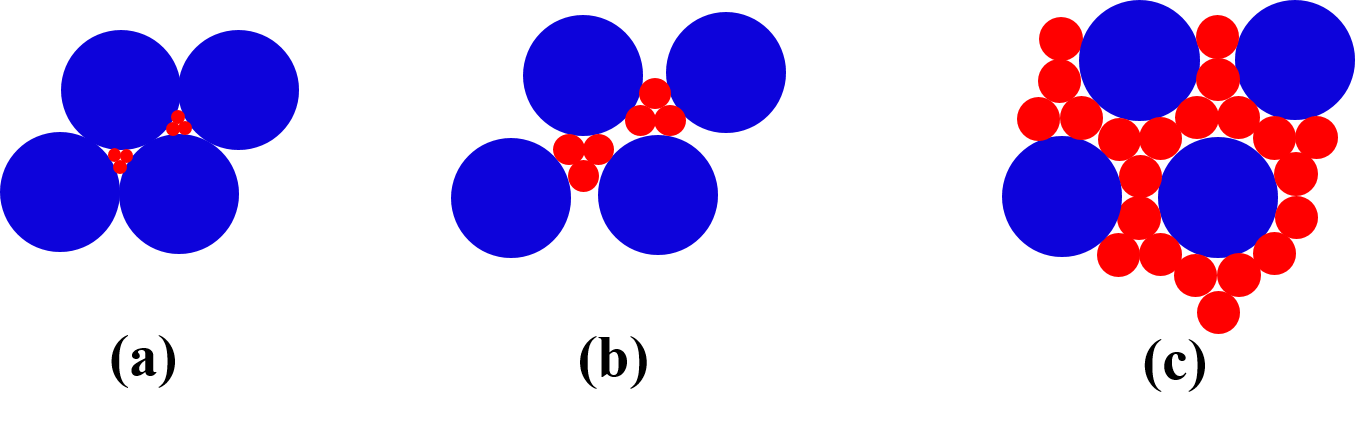}
    \caption{\textbf{Complex T2 motifs with varying microsphere size ratio.} Examples shown with additive microsphere radii.
    (a) Radius ratio $\tilde{a}_\text{S}=0.101$ is the smallest where three-particle arrangements fill the interstitial spaces. (b) Intermediate radius ratio where the large spheres are not in contact. (c) Radius ratio $\tilde{a}_\text{S}=0.394$ where small particle triangles can surround the large ones while maintaining the hexagonal large-particle configuration.}
    \label{fig:placeholder}
\end{figure}

\end{document}